\documentclass[a4paper,12pt]{article}
\usepackage[pdftex]{graphicx}    

\usepackage{geometry}
\usepackage{enumerate}

\usepackage{hyperref}
\usepackage{subcaption}
\usepackage{appendix}
\usepackage{latexsym,booktabs}
\usepackage{amsmath, amssymb, amsthm}
\usepackage{graphicx}
\usepackage{graphics}
\usepackage[singlespacing]{setspace}
\usepackage{color}
\usepackage{tikz}
\usepackage{caption}
\usepackage{subcaption}
\usepackage{booktabs}
\usepackage[pdftex]{graphicx}
\usepackage{geometry}
\usepackage{amsmath, amssymb}
\usepackage{titlesec}
\usepackage{setspace}
\usepackage{mathtools}
\usepackage{amsthm}

\usepackage{algorithm}
\usepackage{algpseudocode}
\theoremstyle{plain}

\titlespacing*{\section}{0pt}{1ex plus .2ex minus .2ex}{1ex plus .2ex}

\title{\textbf{Risk Management with Feature-Enriched Generative Adversarial Networks (FE-GAN)}}
\author{ Ling Chen \\
Department of Mathematics \\
University College London \\
l.chen.23@ucl.ac.uk}
\date{}

\makeatletter
\newcommand{\customlabel}[2]{%
   \protected@write \@auxout {}{\string \newlabel {#1}{{#2}{\thepage}{#2}{#1}{}} }%
   \hypertarget{#1}{#2}
}
\makeatother
\usepackage{cleveref}
\crefname{table}{Table}{Tables}

\begin{document}

% Title and abstract on the same page
\maketitle

\vspace{-1cm} % Adjust spacing after title

\begin{center}
\large{\textbf{Abstract}}
\end{center}

\small
This paper investigates the application of Feature-Enriched Generative Adversarial Networks (FE-GAN) in financial risk management, with a focus on improving the estimation of Value at Risk (VaR) and Expected Shortfall (ES). FE-GAN enhances existing GANs architectures by incorporating an additional input sequence derived from preceding data to improve model performance. Two specialized GANs models, the Wasserstein Generative Adversarial Network (WGAN) and the Tail Generative Adversarial Network (Tail-GAN), were evaluated under the FE-GAN framework. The results demonstrate that FE-GAN significantly outperforms traditional architectures in both VaR and ES estimation. Tail-GAN, leveraging its task-specific loss function, consistently outperforms WGAN in ES estimation, while both models exhibit similar performance in VaR estimation. Despite these promising results, the study acknowledges limitations, including reliance on highly correlated temporal data and restricted applicability to other domains. Future research directions include exploring alternative input generation methods, dynamic forecasting models, and advanced neural network architectures to further enhance GANs-based financial risk estimation.

\vspace{0.2cm} % Add space before keywords

\section{Introduction}

\subsection{Motivation}
Generative Adversarial Networks (GANs), introduced by \cite{goodfellow2014generative}, have achieved remarkable success in machine learning by enabling the generation of realistic data through an adversarial training framework. A GANs architecture comprises two competing networks: a generator and a discriminator. The generator aims to produce data indistinguishable from real samples, while the discriminator seeks to differentiate between real and synthetic data. This adversarial setup iteratively refines the generator, resulting in increasingly realistic outputs.

GANs have found widespread applications across domains, including image synthesis \cite{radford2015unsupervised}, video generation, and text modeling \cite{ledig2017photo}. In financial risk management, GANs show promise for improving estimates of risk measures like Value at Risk (VaR) and Expected Shortfall (ES) by generating synthetic datasets that capture complex financial behaviors. This capability enables robust stress testing and scenario analysis, which are critical for understanding potential risks in volatile markets.

However, despite their utility, GANs often struggle with capturing intricate patterns and temporal dependencies inherent in financial time series data. Addressing these challenges requires innovations that enhance the performance of GANs without altering their underlying architectures.

\subsection{Literature Review}
The use of GANs in financial risk management has garnered increasing attention, particularly for improving VaR and ES estimation. This study builds upon two GANs variants: the Wasserstein Generative Adversarial Network (WGAN) \cite{arjovsky2017wasserstein}, which stabilizes training using the Wasserstein distance, and the Tail Generative Adversarial Network (Tail-GAN) \cite{cont2022tailgan}, which incorporates scoring functions to better model tail risks.

WGAN mitigates common issues like mode collapse by reformulating the loss function, enhancing convergence stability and the quality of generated data. Tail-GAN further refines GANs-based risk estimation by targeting distribution tails, optimizing for measures like VaR and ES through specialized loss functions.

Recent advancements in GANs architectures underscore the importance of leveraging additional information to improve performance. For example, \cite{gulrajani2017improved} demonstrated the value of gradient penalties, while Tail-GAN effectively incorporated financial-tailored loss functions. Nonetheless, little research explores the integration of enriched input features derived from classical models, such as time series methods, as a means to enhance GANs-generated outputs.

\subsection{Contribution}
This paper introduces Feature-Enriched GANs (FE-GAN), a novel modification to GANs designed to improve performance when applied to temporal data. FE-GANs enhance the traditional GANs architecture by incorporating additional input sequences derived from preceding data. This approach is particularly effective for temporal datasets such as financial data, but it can also be applied to any domain where the data is temporal and not severely missing.

The additional input sequences include historical data, as well as outputs generated from classical models like the Geometric Brownian Motion (GBM) model and traditional time series models. We also explore an improved time series model that aggregates the seasonal and trend components from a standard time series model with the volatility component derived from the GBM model. By leveraging these varied input methods, FE-GANs provide a more robust foundation for modeling temporal data, allowing the generator to start closer to the true data distribution, thus accelerating convergence.

Experiments on VIX data, representing the anticipated 30-day volatility of the SP500 index from 2014 to 2019, demonstrate that the FE-GAN architecture significantly reduces estimation errors compared to traditional GANs implementations. In particular, FE-GAN achieves a threefold improvement in convergence speed and a substantial reduction in VaR and ES estimation errors. Notably, the modified architecture was applied to both WGAN and Tail-GAN, and results show that Tail-GAN continues to outperform WGAN under the FE-GAN framework, which aligns with the findings of \cite{cont2022tailgan}. This observation indicates that the improvement brought by FE-GAN maintains the advantage of Tail-GAN over WGAN, suggesting that FE-GAN enhances both models without changing their relative performance.

This work demonstrates the potential of combining classical statistical models with GANs to improve performance for temporal data modeling. FE-GANs offer an effective solution for a wide range of applications where temporal dependencies play a crucial role, such as financial data analysis, forecasting, and beyond.

\subsection{Outlines}

In Section \ref{Modify}, we introduce the FE-GAN framework, which enhances existing GANs architectures by incorporating an additional input sequence derived from historical data, GBM models, or time series analysis. To evaluate its performance, we implement several tests using different input sequences: direct historical data, mean and variance under the GBM assumption, time series analysis, and an improved time series model that integrates components from both time series and GBM. Comparisons among these approaches highlight the effectiveness of the enriched input sequences in improving VaR and ES estimation.

In Section \ref{comp}, we compare the performance of Tail-GAN and WGAN under the FE-GAN framework. This evaluation considers various input sequences introduced in Section \ref{Modify}, including historical data, GBM-based inputs, time series analysis, and the hybrid time series-GBM model. The results demonstrate that Tail-GAN consistently outperforms WGAN in ES estimation, leveraging its tailored loss function for extreme quantiles. For VaR estimation, both models exhibit similar performance, indicating that the enriched input sequences benefit both architectures.

\section{FE-GAN: Architecture and Implementation}
\label{Modify}

This section presents FE-GAN, which enhances GANs by incorporating additional input sequences into the generator. These inputs, derived from preceding data, enable the generator to better capture temporal dependencies and start closer to the true data distribution. This approach accelerates convergence and improves the accuracy of risk measure estimation, particularly for VaR and ES, without modifying the underlying architecture.

Various methods for generating input sequences were explored to demonstrate the effectiveness of FE-GAN:

1. \textbf{Historical Data}: 
   The simplest method uses the historical data immediately preceding the target period. For example, to estimate risk measures for the last two weeks, the generator uses the prior two weeks of data. This method provides substantial performance gains by ensuring that the generator starts training with relevant context.

2. \textbf{GBM Assumption}: 
   This method assumes a Geometric Brownian Motion (GBM) model, generating input sequences based on the mean and variance of past data. Although computationally efficient, it performs less effectively than historical data in estimating VaR and ES for volatility datasets, indicating that GBM does not capture the full dynamics of financial volatility.

3. \textbf{Time Series Analysis}:  
   Time series models, incorporating trend and seasonal components, were also tested. While effective for short-term predictions, their performance diminishes for longer datasets, such as the one-year period used here. Time series methods excel at ES estimation due to their ability to capture tail behavior but show limited improvement in VaR estimation due to a lack of volatility modeling.

4. \textbf{Combined Time Series and GBM Approach}: 
   This hybrid method integrates time series analysis with the volatility term from GBM. The data is decomposed into trend, seasonal, and residual components, with volatility replacing the residual term. This approach improves VaR estimation while maintaining strong ES performance, capturing both long-term trends and short-term volatility.
   
The results demonstrate that FE-GAN consistently outperforms traditional WGAN, with the choice of input sequence influencing the magnitude of improvement. The purpose of testing with raw historical data, a simple GBM model, and a traditional time series model (which is more complex than GBM) is to assess FE-GAN's sensitivity to different levels of prediction models. In other words, we aim to explore how well FE-GAN generalizes when the underlying prediction model is relatively poor. By using three different models of varying complexity, we test whether FE-GAN can adapt to and improve upon less accurate inputs. The final aggregation model combines these diverse approaches to enhance the generator's performance. While this may not represent the best predictive model, it demonstrates the potential of aggregating multiple models to create a stronger, more robust input for the generator.

\subsection{Architecture of FE-GAN}
\label{modified}

\begin{figure}[h!]
    \centering
    \includegraphics[width=\linewidth]{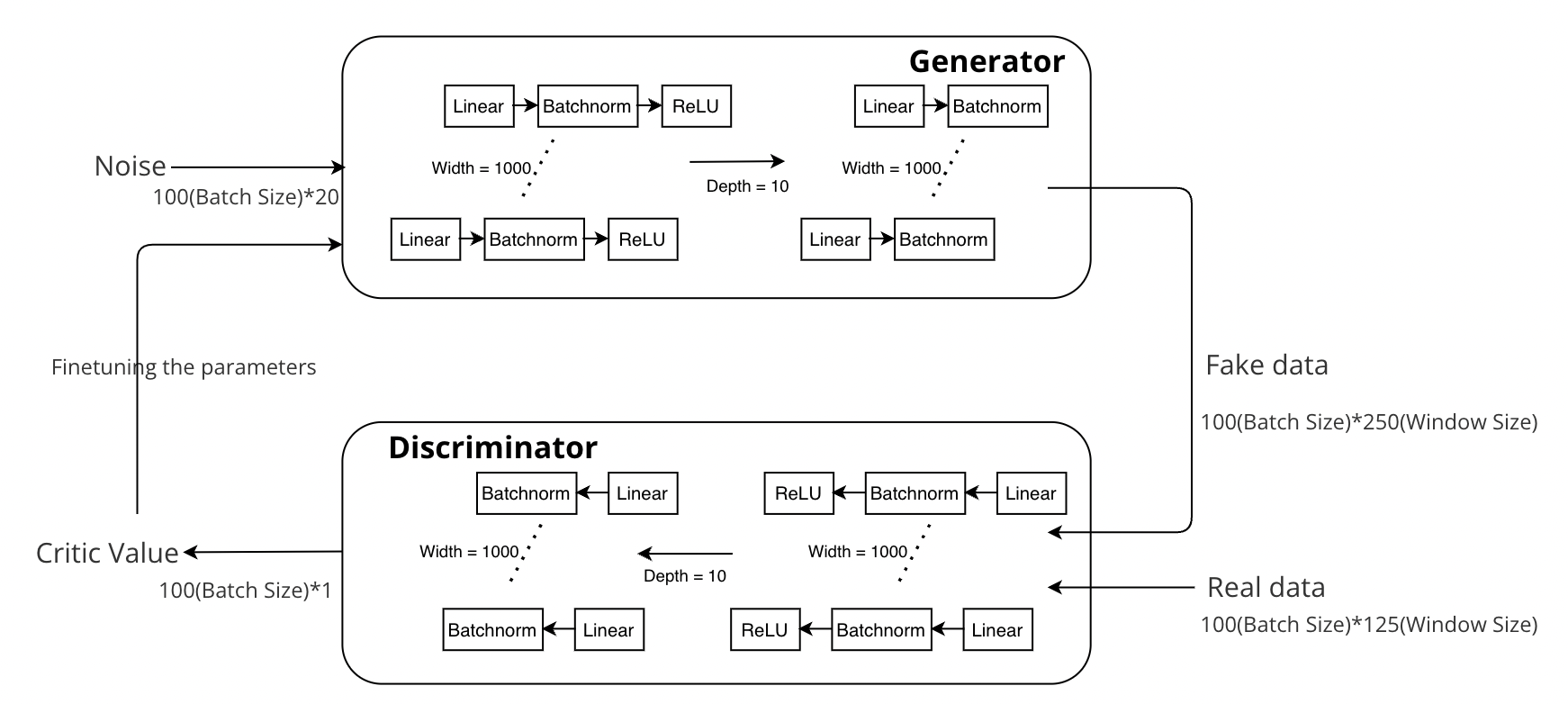}
    \caption{Traditional WGAN architecture.}
    \label{fig:GAN}
\end{figure}

The traditional WGAN architecture consists of two core components: a generator and a discriminator. The generator takes random noise as input and learns to produce synthetic data that closely resembles the real data, with the aim of fooling the discriminator. The discriminator distinguishes between real and generated data, providing feedback to the generator. This adversarial process iteratively refines the generator, improving the realism of the synthetic data. The standard WGAN structure is depicted in Figure \ref{fig:GAN}.

For this work, a linear GANs architecture was implemented, where each layer consists of a linear transformation, batch normalization, and a ReLU activation function. The training configuration includes a batch size of 100 and a window size of 250, corresponding to 250 trading days. During each training iteration, 100 random sequences of length 250 are sampled. The generator consists of 10 layers with a width of 1,000, while the discriminator uses a narrower configuration with 5 layers and a width of 100.

\begin{figure}[h]
    \centering
    \includegraphics[width=\linewidth]{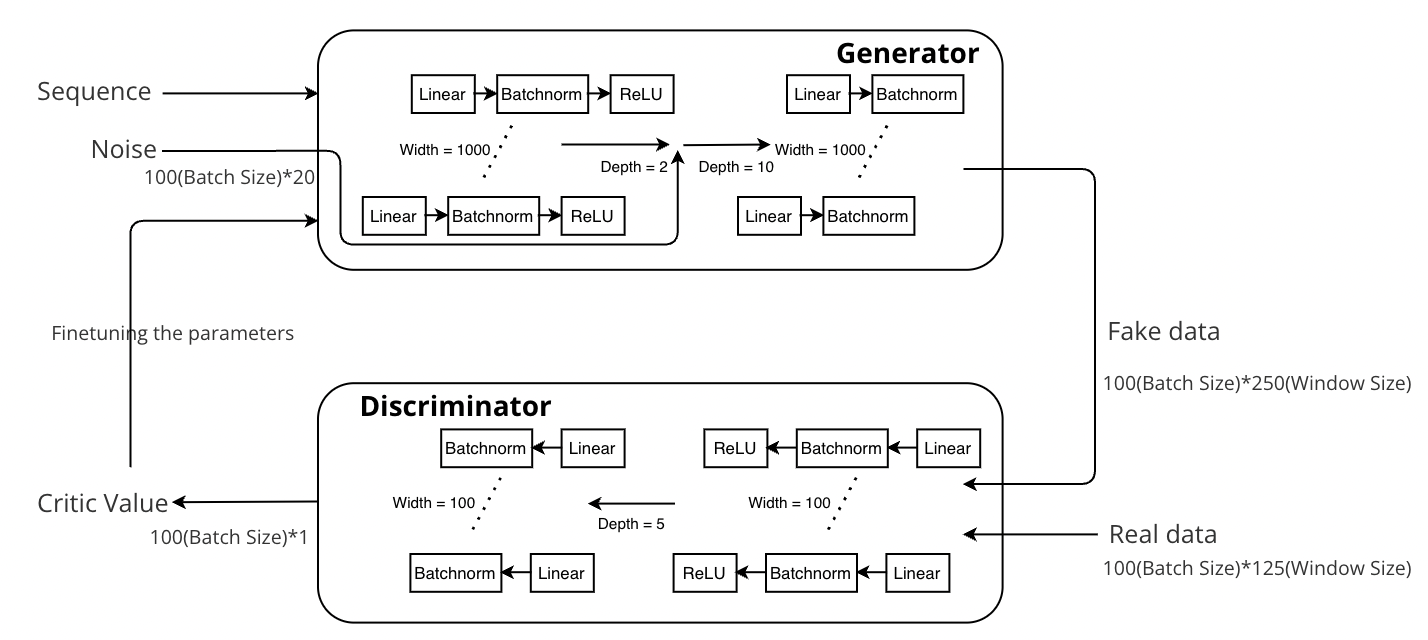}
    \caption{FE-GAN architecture.}
    \label{fig:Modified GAN}
\end{figure}

The enhancement introduced in this work incorporates an additional input sequence into the generator. This input, derived from historical data, provides contextual information that helps the generator start closer to the true data distribution. The structure of this enhanced generator is shown in Figure \ref{fig:Modified GAN}.

In FE-GAN, each training batch includes 100 historical sequences, which are used to predict the next 250 values. The historical data is processed through two preprocessing layers designed to extract key features before being combined with random noise. This combined input is then fed into the generator, which retains the original WGAN structure. This additional input helps the generator converge more quickly to the target distribution by providing it with relevant context from prior data.

A related approach is discussed in \cite{wang2024forecasting}, where specific VaR levels are used directly as inputs for financial risk estimations. However, such methods are typically designed for narrowly defined tasks and lack the flexibility for broader applications. In contrast, the approach presented here introduces flexibility by allowing the generator to adapt to various financial measures, including multiple VaR and ES levels, as well as other divergence metrics such as the Kullback-Leibler divergence. This adaptability ensures robust performance across different financial conditions, making the approach more suitable for practical applications.

\subsection{Test of Historical Data}

The implementation of FE-GAN significantly improves performance, even when using the simplest input data—historical data. By incorporating the preceding 250 data points as input along with random noise, and without altering other hyperparameters, the results demonstrate substantial improvements in risk measure estimation.

\begin{figure}[h]
    \centering
    % First subfigure
    \begin{subfigure}[b]{0.45\textwidth}
        \centering
        \includegraphics[width=\linewidth]{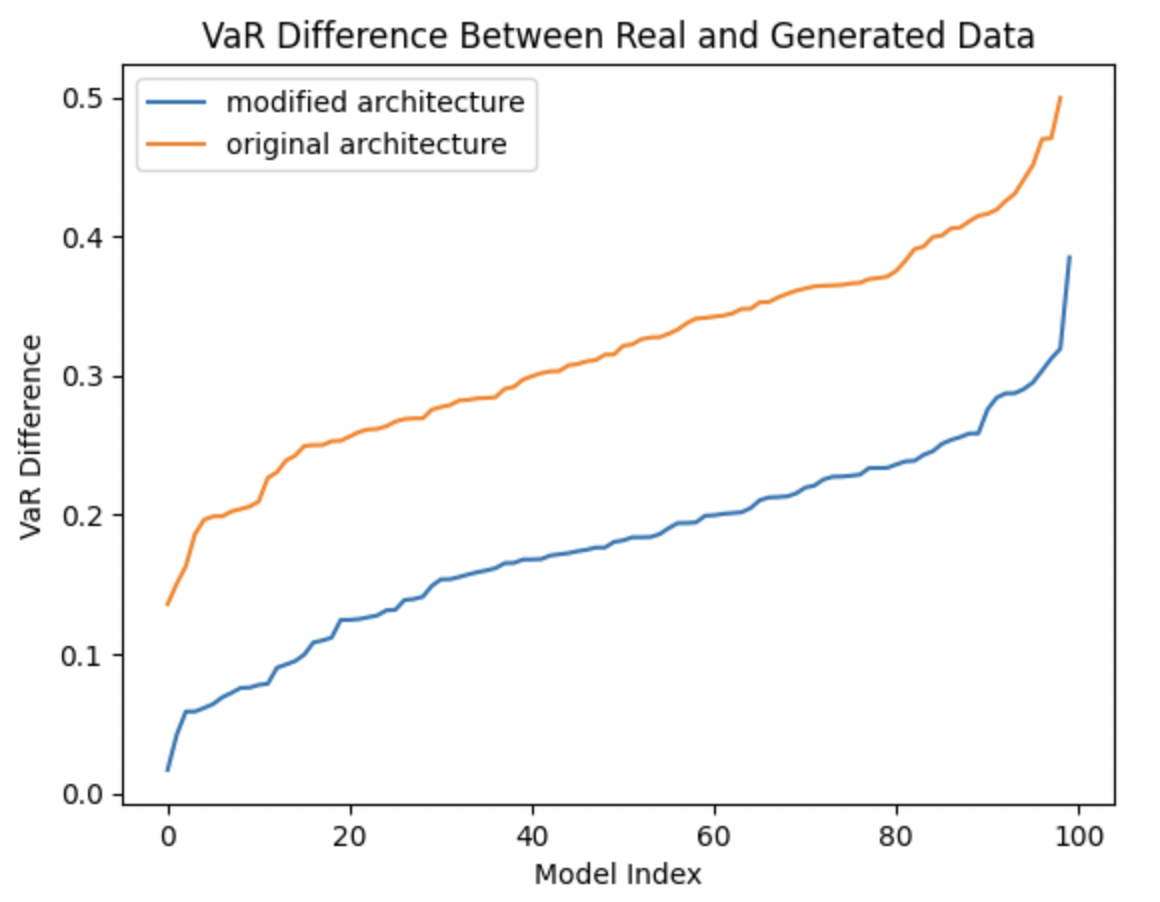}
        \caption{Comparison of VaR difference between the benchmark and the test of historical data.}
        \label{fig:VaR(origin vs hist)}
    \end{subfigure}
    \hfill
    % Second subfigure
    \begin{subfigure}[b]{0.45\textwidth}
        \centering
        \includegraphics[width=\linewidth]{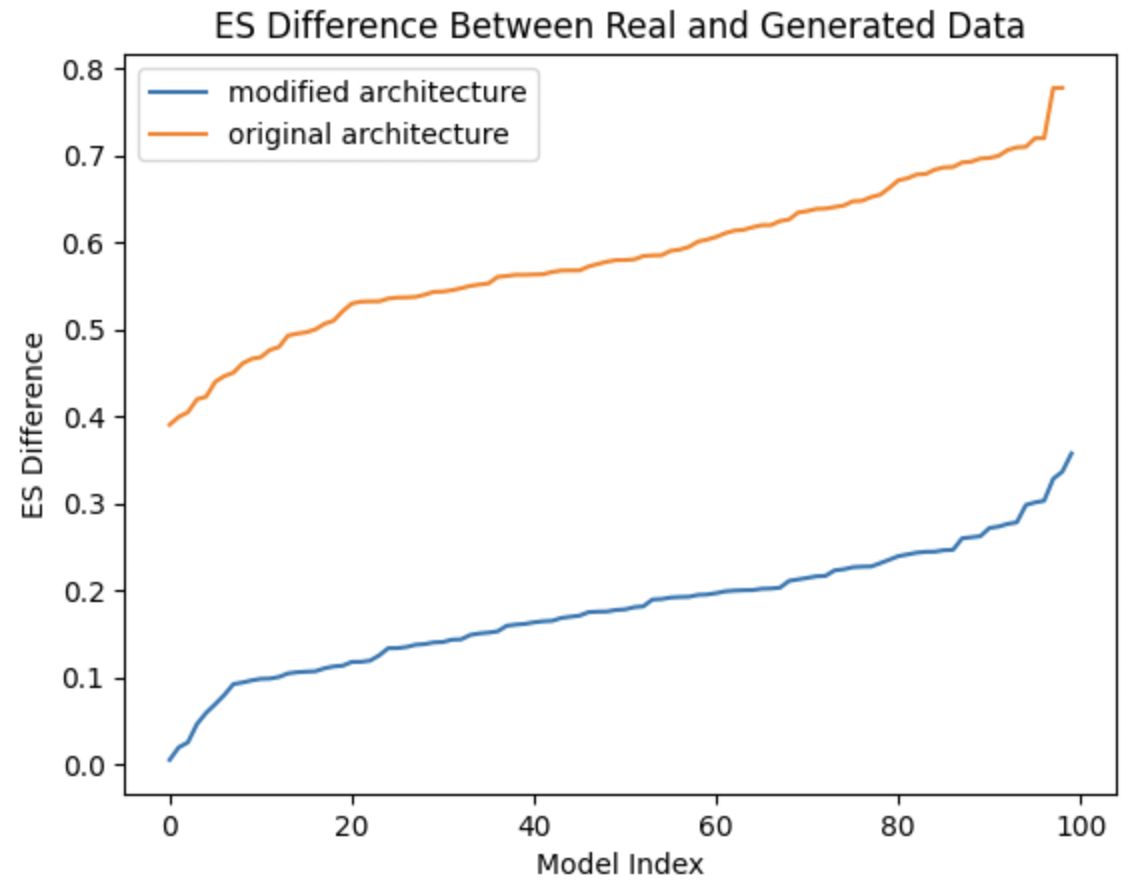}
        \caption{Comparison of ES difference between the benchmark and the test of historical data.}
        \label{fig:ES(origin vs hist)}
    \end{subfigure}
    \caption{Comparison of VaR and ES differences at the 5\% level between the benchmark and the test of historical data.}
    \label{fig:comparison}
\end{figure}

As shown in Figure \ref{fig:VaR(origin vs hist)}, the VaR difference for the benchmark model varies between 0.1 and 0.5 across 100 models, while FE-GAN, using historical data as input, exhibits VaR differences ranging from 0.01 to 0.34, with 90\% of values falling under 0.25. This indicates a marked improvement in estimating VaR at the 5\% level. Furthermore, the training time is reduced to approximately one tenth of the time required by the benchmark model.

Figure \ref{fig:ES(origin vs hist)} shows the ES differences between the two architectures. FE-GAN demonstrates an even more pronounced improvement, with its highest difference being lower than the smallest difference observed in the benchmark model. On average, FE-GAN reduces the loss by more than half when compared to the benchmark.

\subsection{Test under the Assumption of GBM}

In a second test, we assume that the VIX data follows a GBM model. Specifically, the GBM is governed by the stochastic differential equation:
\begin{equation}
dS_t = \mu S_t \, dt + \sigma S_t \, dW_t,
\end{equation}
where \( S_t \) is the VIX value at time \( t \), \( \mu \) is the drift coefficient, \( \sigma \) is the volatility coefficient, and \( W_t \) is a Wiener process.

This implies that the log returns follow a normal distribution:
\begin{equation}
\log \frac{S_t}{S_{t-1}} \sim \mathcal{N} \left( \mu - \frac{\sigma^2}{2}, \sigma^2 \right).
\end{equation}

Under the GBM assumption, the log returns of the data follow a normal distribution, simplifying the implementation. It implies that the values of the cleaned data at any time \( t \) are independent and identically distributed (i.i.d.). While this assumption is a simplification, it offers convenience. Using the historical mean and variance, we can easily simulate a sequence of length 250 that follows a normal distribution with these parameters.

\begin{figure}[h]
    \centering
    % First subfigure
    \begin{subfigure}[b]{0.45\textwidth}
        \centering
        \includegraphics[width=\linewidth]{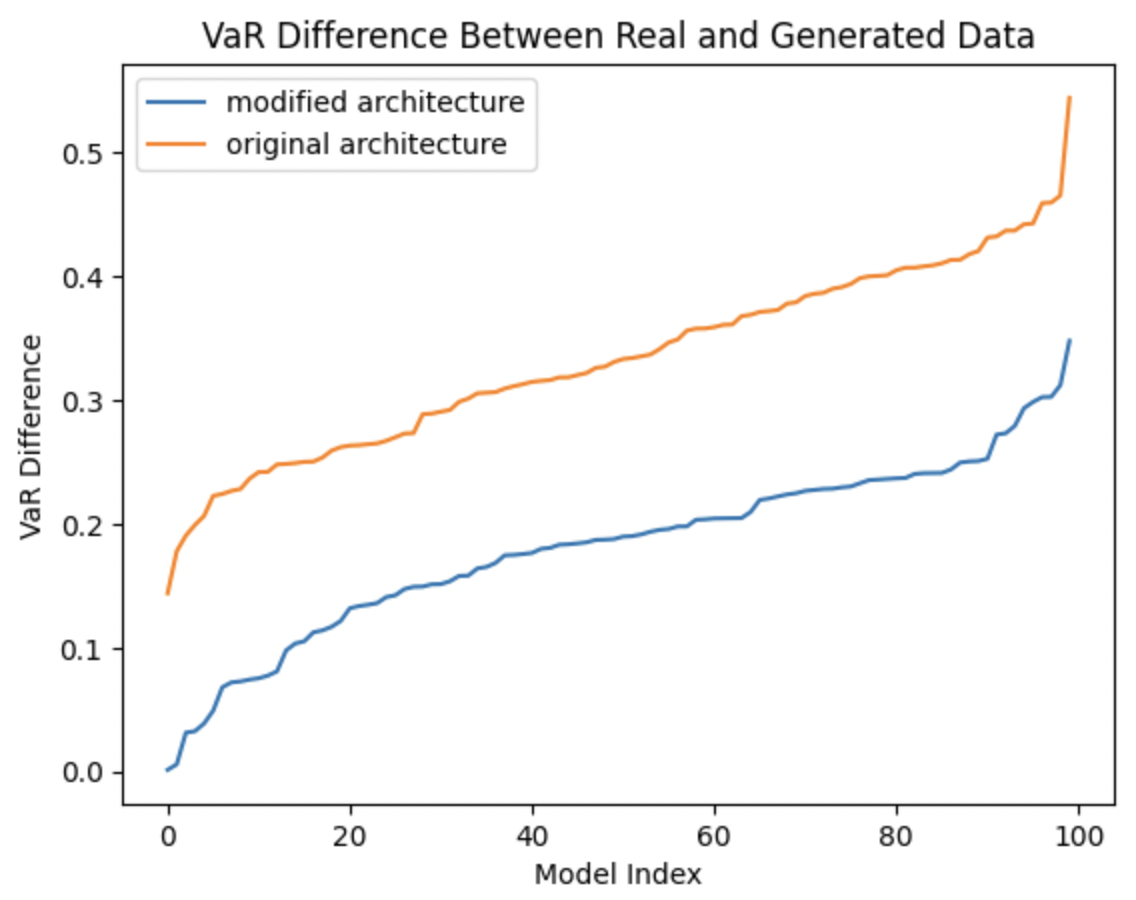}
        \caption{Comparison of VaR difference between the benchmark and the test under the GBM assumption.}
        \label{fig:VaR(origin vs GBM)}
    \end{subfigure}
    \hfill
    % Second subfigure
    \begin{subfigure}[b]{0.45\textwidth}
        \centering
        \includegraphics[width=\linewidth]{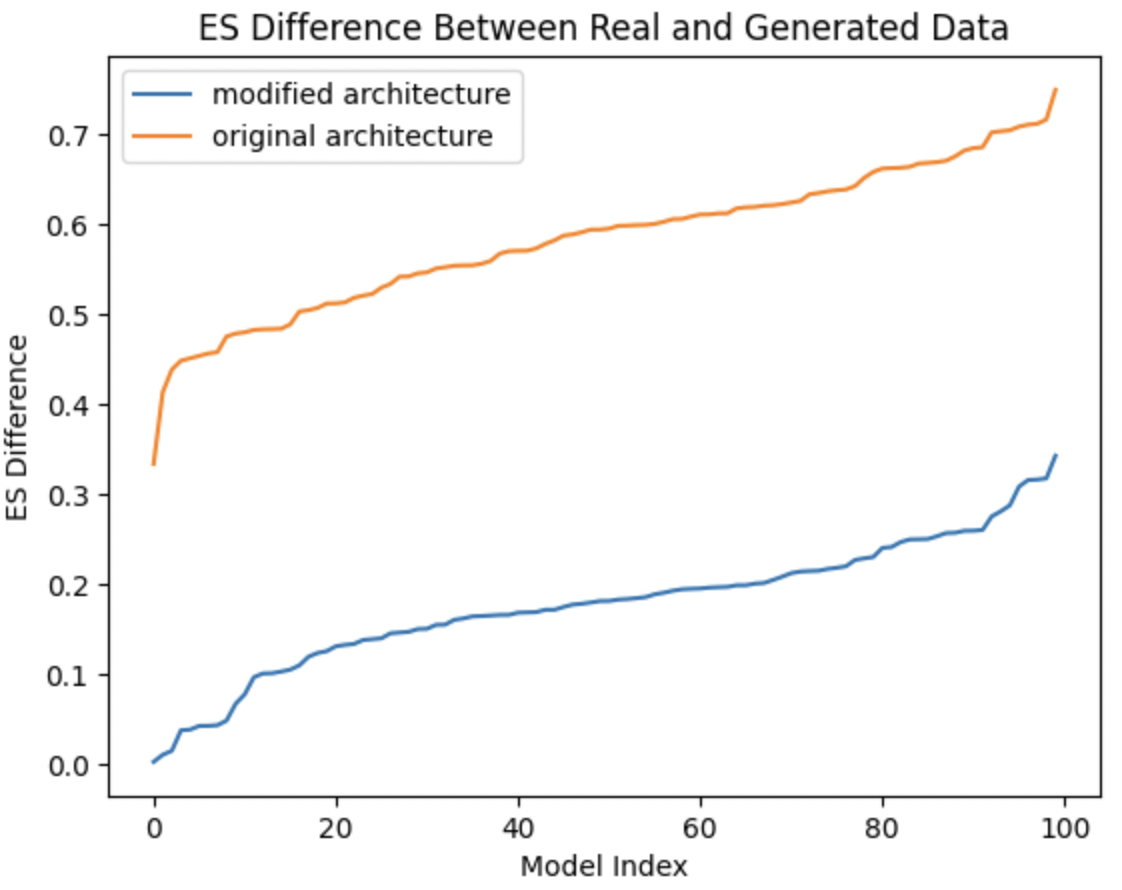}
        \caption{Comparison of ES difference between the benchmark and the test under the GBM assumption.}
        \label{fig:ES(origin vs GBM)}
    \end{subfigure}
    \caption{Comparison of VaR and ES differences at the 5\% level between the benchmark and the test under the GBM assumption.}
    \label{fig:comparison_GBM}
\end{figure}

Figure \ref{fig:VaR(origin vs GBM)} illustrates that the VaR difference under the GBM assumption varies from 0.02 to 0.35, showing better performance than the benchmark model, and similar to the first test using historical data (see Figure \ref{fig:VaR(origin vs hist)}). Similarly, Figure \ref{fig:ES(origin vs GBM)} shows that the ES difference varies from 0.01 to 0.35, also comparable to the historical data test (see Figure \ref{fig:ES(origin vs hist)}). These results suggest that using the mean and variance under the GBM assumption yields performance similar to using historical data, with both approaches requiring similar training times. This indicates that both methods capture similar levels of trend and volatility present in the true data.

\subsection{Test of Time Series}

In time series analysis, selecting the optimal model is crucial for balancing prediction accuracy and avoiding overfitting. The Akaike Information Criterion (AIC) and Bayesian Information Criterion (BIC) are widely used tools in this regard, providing a systematic approach to model selection by evaluating both the goodness-of-fit and the complexity of the model. These criteria help ensure that models are neither too simple to underfit the data nor too complex to overfit, particularly when working with limited data where overfitting can be a significant risk.

AIC is commonly preferred in predictive modeling tasks, as it prioritizes minimizing prediction error. This makes AIC particularly suitable when the goal is to develop a model that generalizes well to new data, as it offers more flexibility in selecting complex models that can better capture the underlying data patterns. In the context of time series forecasting, where underlying structures such as seasonality, autocorrelation, and volatility clustering often exist, AIC allows for a more nuanced selection of models, particularly when capturing such dynamics requires a greater number of parameters \cite{akaike1974new}.

In contrast, BIC places a stronger emphasis on model simplicity by applying a more substantial penalty for the number of parameters, particularly as sample sizes increase. This makes BIC a useful tool when there is a concern that more complex models might overfit, especially in large datasets where the risk of overfitting increases. BIC tends to favor simpler models that may not capture all the nuances of the data but are less likely to become overly complex and less generalizable \cite{schwarz1978estimating}. Therefore, while BIC is conservative, it can serve as an important safeguard in model selection, especially when the primary concern is avoiding unnecessary complexity.

We use rolling windows of the dataset as predictive sequences, testing each model with the preceding 500 data points. If a model performs well across most rolling windows, it is assumed to generalize well to the entire dataset. To identify the most appropriate models, we tested autoregressive (AR) and moving average (MA) models up to the third order, as well as ARMA models up to the second order. For each model, we calculated the AIC and BIC values, which are summarized in \hyperref[table:acf_pacf_part1]{Table 1.1}, \hyperref[table:acf_pacf_part2]{Table 1.2}, and \hyperref[table:acf_pacf_part3]{Table 1.3}. These tables present the AIC and BIC values for each model across different rolling windows of the data.

The AIC and BIC analyses suggest that the AR(1) model performs best among the AR models, while the MA(1) model performs best among the MA models, yielding the smallest AIC and BIC values. Among the ARMA models, both ARMA(1,2) and ARMA(2,1) demonstrate comparable performance, consistently outperforming other models. Compared to the AR(1) and MA(1) models, ARMA(1,2) and ARMA(2,1) show superior overall performance. Based on these findings, we select the ARMA(2,1) model for all sequence generation within our improved WGAN architecture. This model provides a good trade-off between predictive accuracy and model simplicity, ensuring both accurate predictions and faster convergence in the modeling process.

In this study, AIC is prioritized due to its focus on predictive accuracy. Given that we are working with rolling windows and repeatedly testing models across different sequences of data, AIC’s flexibility in selecting more complex models is advantageous for capturing the full range of temporal dependencies present in the data. This is particularly relevant in financial time series data, where patterns such as volatility clustering and non-stationarity are common and require more adaptable models \cite{hamilton2020time}. However, BIC is also considered for comparison to ensure that the models selected do not become overly complex.

\begin{table}[h!]
\centering
\caption{AIC and BIC Values for Different Models}
\label{table:main}

\captionsetup[subtable]{labelformat=simple, labelsep=colon}
\renewcommand{\thesubtable}{Table 1.\arabic{subtable}}

\begin{subtable}[t]{{1\textwidth}}
\centering
\caption{AIC and BIC Values for Different Models (Part 1)}
\label{table:acf_pacf_part1}
\resizebox{\textwidth}{!}{%
\begin{tabular}{|c|cc|cc|cc|cc|}
\hline
\textbf{Model} & \multicolumn{2}{c|}{\textbf{1-500}} & \multicolumn{2}{c|}{\textbf{51-550}} & \multicolumn{2}{c|}{\textbf{101-600}} & \multicolumn{2}{c|}{\textbf{151-650}} \\
 & \textbf{AIC} & \textbf{BIC} & \textbf{AIC} & \textbf{BIC} & \textbf{AIC} & \textbf{BIC} & \textbf{AIC} & \textbf{BIC} \\
\hline
\textbf{AR(1)} & -1215.99 & -1203.352 & -1206.528 & -1193.896 & -1185.193 & -1172.562 & -1182.874 & -1170.243 \\
\textbf{AR(2)} & -1211.318 & -1194.476 & -1201.308 & -1184.474 & -1180.649 & -1163.815 & -1179.009 & -1162.174 \\
\textbf{AR(3)} & -1206.103 & -1185.06 & -1196.378 & -1175.345 & -1175.417 & -1154.384 & -1173.945 & -1152.912 \\
\textbf{MA(1)} & -1218.861 & -1206.217 & -1209.813 & -1197.175 & -1188.585 & -1175.947 & -1186.355 & -1173.717 \\
\textbf{MA(2)} & -1217.738 & -1200.879 & -1208.001 & -1191.151 & -1187.529 & -1170.678 & -1185.447 & -1168.596 \\
\textbf{MA(3)} & -1216.004 & -1194.931 & -1206.158 & -1185.095 & -1185.819 & -1164.756 & -1183.74 & -1162.677 \\
\textbf{ARMA(1,1)} & -1217.7 & -1200.842 & -1207.793 & -1190.943 & -1187.055 & -1170.204 & -1185.256 & -1168.405 \\
\textbf{ARMA(1,2)} & -1231.83 & -1210.757 & -1206.003 & -1184.94 & -1201.055 & -1179.992 & -1200.711 & -1179.647 \\
\textbf{ARMA(2,1)} & -1231.825 & -1210.752 & -1221.074 & -1200.011 & -1201.029 & -1179.966 & -1200.65 & -1179.587 \\
\textbf{ARMA(2,2)} & -1230.431 & -1205.143 & -1219.639 & -1194.363 & -1199.154 & -1173.878 & -1197.776 & -1172.5 \\
\hline
\end{tabular}%
}
\end{subtable}

\vspace{0.5cm}

\begin{subtable}[t]{1\textwidth}
\centering
\caption{AIC and BIC Values for Different Models (Part 2)}
\label{table:acf_pacf_part2}
\resizebox{\textwidth}{!}{%
\begin{tabular}{|c|cc|cc|cc|cc|}
\hline
\textbf{Model} & \multicolumn{2}{c|}{\textbf{201-700}} & \multicolumn{2}{c|}{\textbf{251-750}} & \multicolumn{2}{c|}{\textbf{301-800}} & \multicolumn{2}{c|}{\textbf{351-850}} \\
 & \textbf{AIC} & \textbf{BIC} & \textbf{AIC} & \textbf{BIC} & \textbf{AIC} & \textbf{BIC} & \textbf{AIC} & \textbf{BIC} \\
\hline
\textbf{AR(1)} & -1200.908 & -1188.276 & -1224.944 & -1212.312 & -1201.644 & -1189.012 & -1235.385 & -1222.753 \\
\textbf{AR(2)} & -1196.108 & -1179.273 & -1220.805 & -1203.971 & -1198.453 & -1181.618 & -1235.316 & -1218.481 \\
\textbf{AR(3)} & -1191.506 & -1170.473 & -1216.133 & -1195.1 & -1193.799 & -1172.766 & -1233.271 & -1212.238 \\
\textbf{MA(1)} & -1203.62 & -1190.982 & -1227.913 & -1215.275 & -1205.038 & -1192.4 & -1209.356 & -1196.718 \\
\textbf{MA(2)} & -1202.215 & -1185.364 & -1227.09 & -1210.24 & -1204.167 & -1187.316 & -1209.168 & -1192.318 \\
\textbf{MA(3)} & -1200.539 & -1179.476 & -1226.081 & -1205.018 & -1202.973 & -1181.91 & -1209.139 & -1188.076 \\
\textbf{ARMA(1,1)} & -1202.224 & -1185.374 & -1226.484 & -1209.633 & -1203.657 & -1186.806 & -1215.428 & -1198.577 \\
\textbf{ARMA(1,2)} & -1215.4 & -1194.337 & -1241.038 & -1219.975 & -1215.59 & -1194.527 & -1215.908 & -1194.845 \\
\textbf{ARMA(2,1)} & -1215.416 & -1194.352 & -1240.946 & -1219.883 & -1215.484 & -1194.421 & -1213.278 & -1192.215 \\
\textbf{ARMA(2,2)} & -1212.979 & -1187.704 & -1236.672 & -1211.396 & -1211.61 & -1186.334 & -1213.352 & -1188.076 \\
\hline
\end{tabular}%
}
\end{subtable}

\vspace{0.5cm}

\begin{subtable}[t]{1\textwidth}
\centering
\caption{AIC and BIC Values for Different Models (Part 3)}
\label{table:acf_pacf_part3}
\resizebox{\textwidth}{!}{%
\begin{tabular}{|c|cc|cc|cc|cc|}
\hline
\textbf{Model} & \multicolumn{2}{c|}{\textbf{401-900}} & \multicolumn{2}{c|}{\textbf{451-950}} & \multicolumn{2}{c|}{\textbf{501-1000}} & \multicolumn{2}{c|}{\textbf{501-1050}} \\
 & \textbf{AIC} & \textbf{BIC} & \textbf{AIC} & \textbf{BIC} & \textbf{AIC} & \textbf{BIC} & \textbf{AIC} & \textbf{BIC} \\
\hline
\textbf{AR(1)} & -1260.152 & -1247.52 & -1315.696 & -1303.064 & -1252.323 & -1239.691 & -1275.412 & -1262.78 \\
\textbf{AR(2)} & -1255.39 & -1238.556 & -1313.545 & -1296.711 & -1249.34 & -1232.506 & -1272.412 & -1255.578 \\
\textbf{AR(3)} & -1251.498 & -1230.465 & -1308.013 & -1286.98 & -1243.951 & -1222.918 & -1268.333 & -1247.3 \\
\textbf{MA(1)} & -1263.776 & -1251.138 & -1168.991 & -1156.67 & -1254.143 & -1241.505 & -1277.212 & -1264.574 \\
\textbf{MA(2)} & -1262.44 & -1245.59 & -1167.008 & -1150.58 & -1254.628 & -1237.778 & -1277.698 & -1260.848 \\
\textbf{MA(3)} & -1261.23 & -1240.167 & -1165.575 & -1145.04 & -1252.658 & -1231.595 & -1275.728 & -1254.665 \\
\textbf{ARMA(1,1)} & -1269.437 & -1252.587 & -1316.817 & -1299.966 & -1266.987 & -1250.137 & -1290.056 & -1273.205 \\
\textbf{ARMA(1,2)} & -1267.754 & -1246.691 & -1314.866 & -1293.803 & -1265.053 & -1243.99 & -1288.122 & -1267.059 \\
\textbf{ARMA(2,1)} & -1267.755 & -1246.692 & -1314.794 & -1293.731 & -1252.387 & -1231.324 & -1275.456 & -1254.393 \\
\textbf{ARMA(2,2)} & -1265.443 & -1240.167 & -1319.986 & -1294.711 & -1263.377 & -1238.101 & -1286.446 & -1261.17 \\
\hline
\end{tabular}%
}
\end{subtable}
\end{table}

\begin{figure}[h]
    \centering
    \begin{subfigure}[b]{0.45\textwidth}
        \centering
        \includegraphics[width=\linewidth]{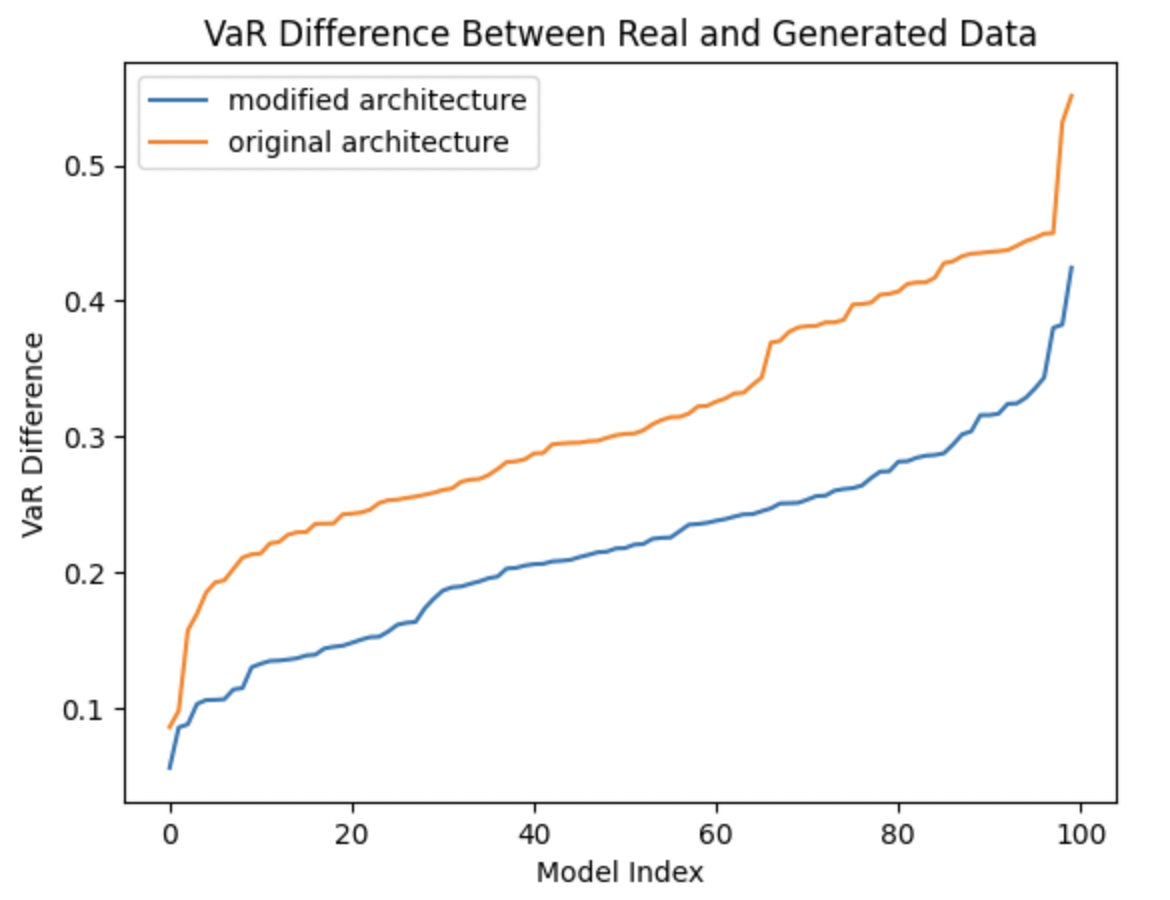}
        \caption{Comparison of VaR difference between the benchmark and the trial of time series}
        \label{fig:VaR(origin vs time series)}
    \end{subfigure}
    \hfill
    \begin{subfigure}[b]{0.45\textwidth}
        \centering
        \includegraphics[width=\linewidth]{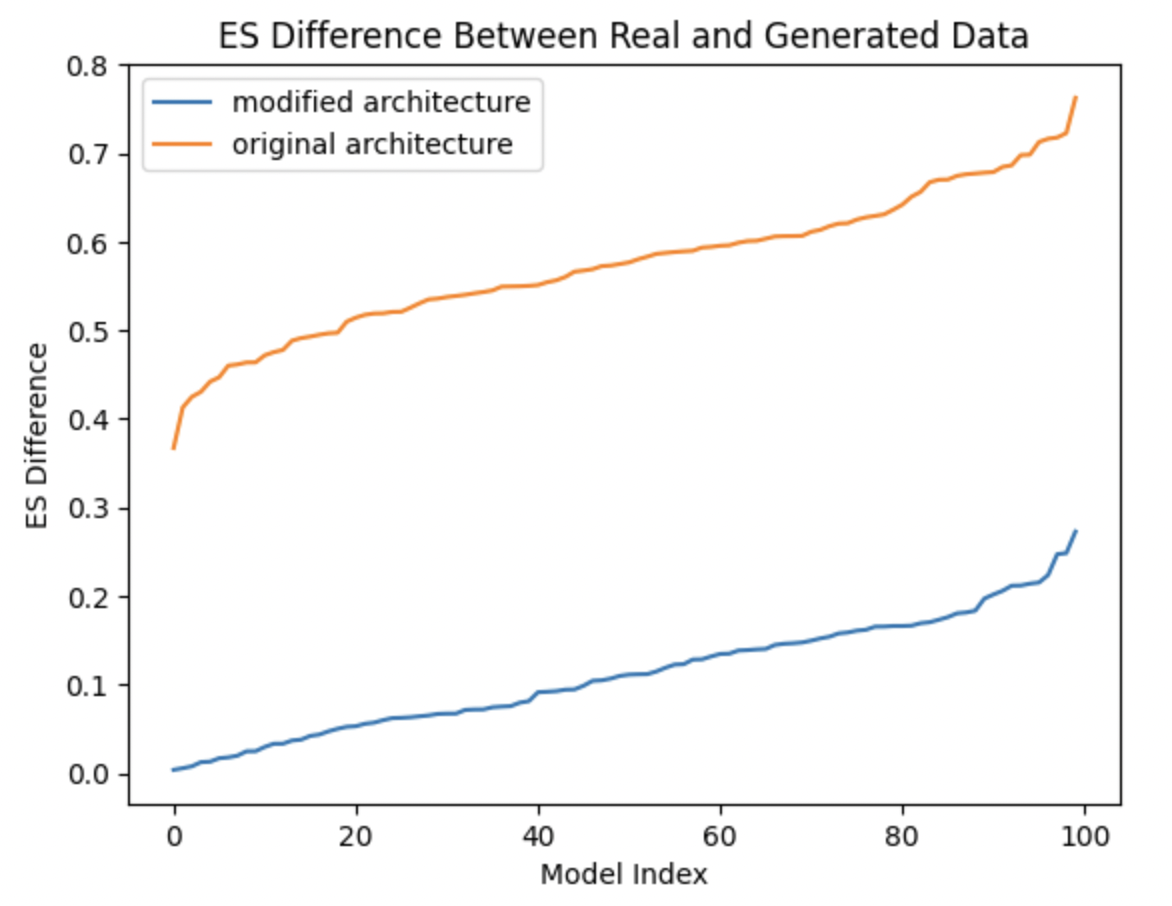}
        \caption{Comparison of ES difference between the benchmark and the trial of time series}
        \label{fig:ES(origin vs time series)}
    \end{subfigure}
    
    \caption{Comparison of VaR and ES differences at 5\% level between the benchmark and the time series model}
\end{figure}

Figure \ref{fig:VaR(origin vs time series)} shows a slightly worse performance than the previous two trials, where the median VaR difference is around 0.21, compared to 0.17 in both previous trials (see Figures \ref{fig:VaR(origin vs hist)} and \ref{fig:VaR(origin vs GBM)}), and the highest difference exceeds 0.35. However, Figure \ref{fig:ES(origin vs time series)} indicates that the ES difference is reduced significantly, with 99 out of 100 models under 0.22, and the median around 0.12, whereas the previous two trials have a median around 0.2 and the highest difference exceeding 0.35 (see Figures \ref{fig:ES(origin vs hist)} and \ref{fig:ES(origin vs GBM)}). This dramatic improvement, approximately 40\% better, is largely attributed to the capability of time series models in capturing trend features. It is evident to perform even better for shorter time periods.

\begin{figure}[h!]
    \centering
    \includegraphics[width=0.5\linewidth]{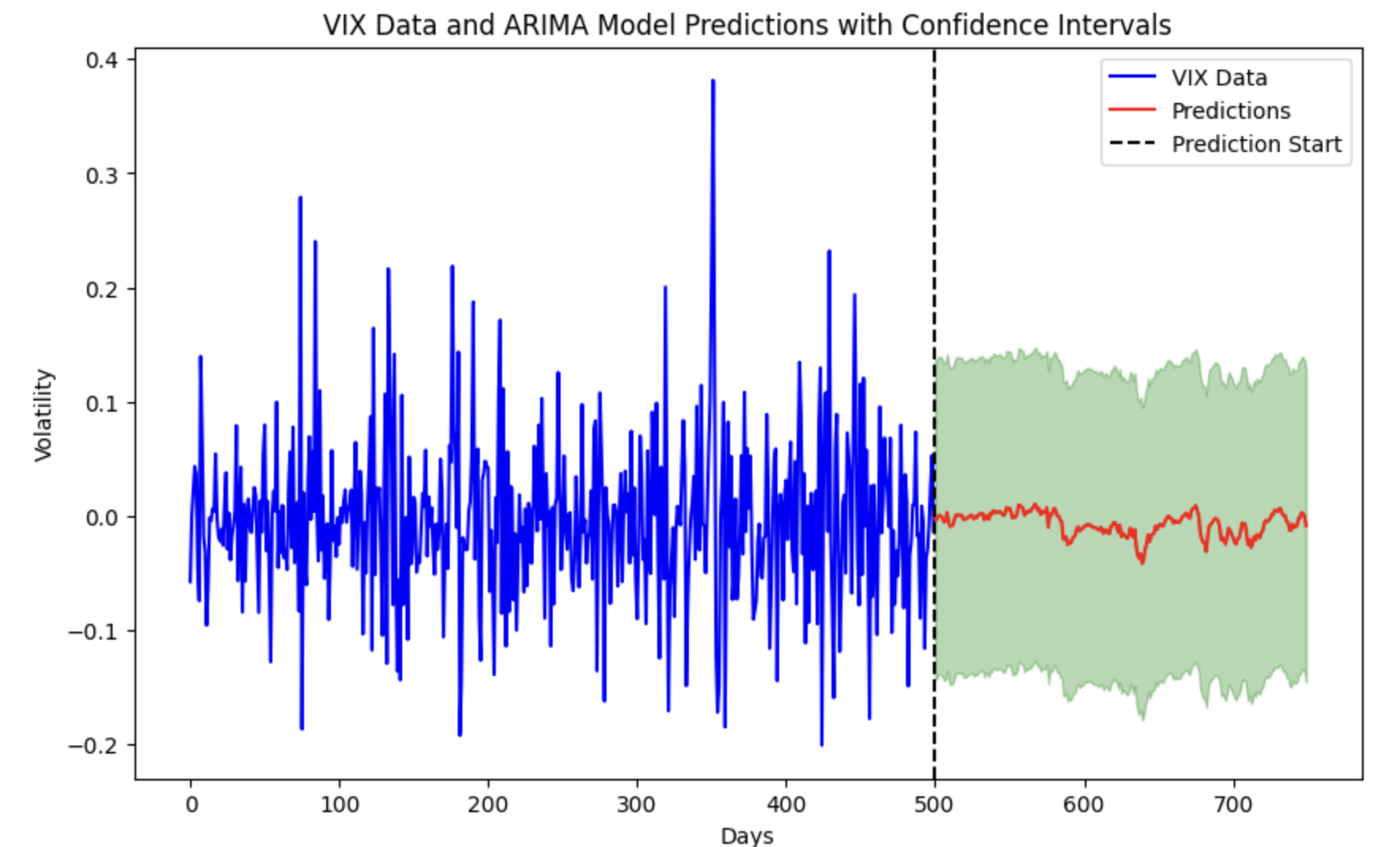}
    \caption{Prediction for 250 data points using the preceding 500 data points(first 500 data points in cleaned VIX) by ARMA(2,1), where the green area is 95\% confidence interval}

    \label{fig:Prediction}
\end{figure}

To visualize that time series analysis has a weaker ability to capture volatility features and a stronger ability to forecast trends, let us use the example from the previous section(using the first 500 data points to predict the 501-750 data points) and compare both the historical data and predicted data. Figure \ref{fig:Prediction} demonstrates that while the prediction captures the main trend effectively, it fails to adequately reflect the volatility, even over a short period. This bad prediction is due to several factors. First, trends are typically more stable and predictable, making them easier to model. In contrast, volatility is inherently random and prone to sudden spikes, which are harder to predict. Second, many time series models, including the ARMA model used here, prioritize capturing linear relationships and assume constant variance (homoscedasticity). These models are designed to smooth out fluctuations to detect the underlying trend, which can cause them to overlook or inadequately capture the dynamic nature of volatility.

To improve VaR estimation while preserving the advantages of ES estimation, we propose decomposing the time series into its trend, noise, and seasonal components. This approach leverages time series analysis for accurate trend prediction and the GBM assumption for modeling volatility. By combining the trend and seasonal components derived from time series analysis with the volatility modeled under the GBM assumption, this hybrid method aims to enhance the accuracy of VaR estimates. The following section will detail the decomposition process and outline the subsequent training methodology.

\subsection{Combination of GBM and Time Series Models}

We aim to leverage the strengths of both the GBM and time series models while mitigating their respective weaknesses. The GBM model effectively captures volatility but performs poorly in trend prediction, leading to strong VaR estimations but weaker ES estimations. Conversely, time series models, while less accurate for VaR, significantly enhance trend prediction, improving ES estimations by approximately 40\%.

Given the decomposition of the time series into trend, seasonal, and noise components, we isolate the volatility component from the GBM model (with the mean set to zero) and combine it with the trend and seasonal components from time series analysis. Using the additive model, which assumes the observed value is the sum of these components, we replace the noise component with the variance derived from historical data. This hybrid model improves both VaR and ES estimations.

\begin{figure}[h]
    \begin{minipage}{0.4\linewidth}
        \centering
        \includegraphics[width=\linewidth]{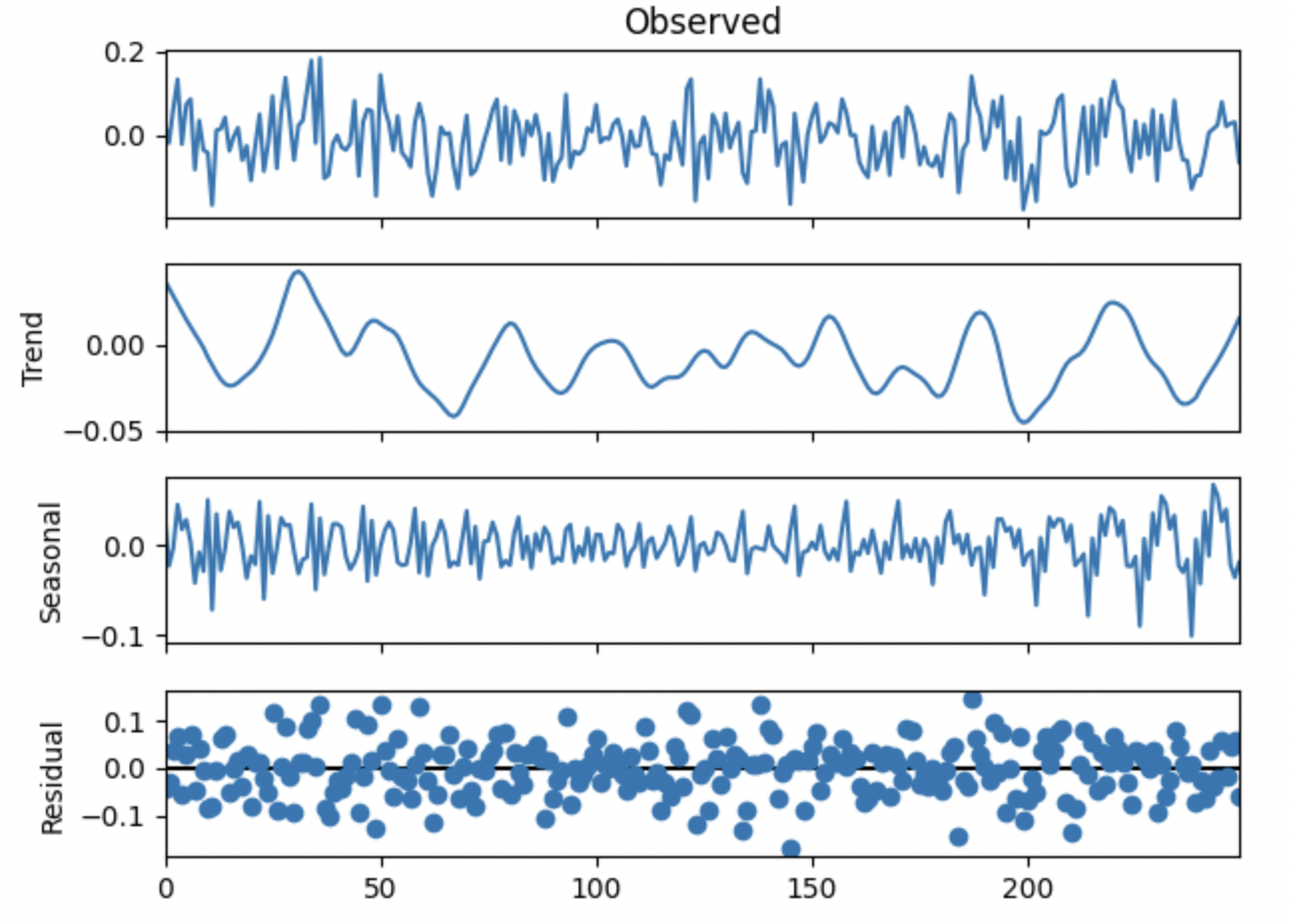}
        \caption{Decomposition of the predicted 250 data points}
        \label{decomposition}
    \end{minipage}
    \hspace{0.1\textwidth}
    \begin{minipage}{0.4\linewidth}
        \centering
        \includegraphics[width=\linewidth]{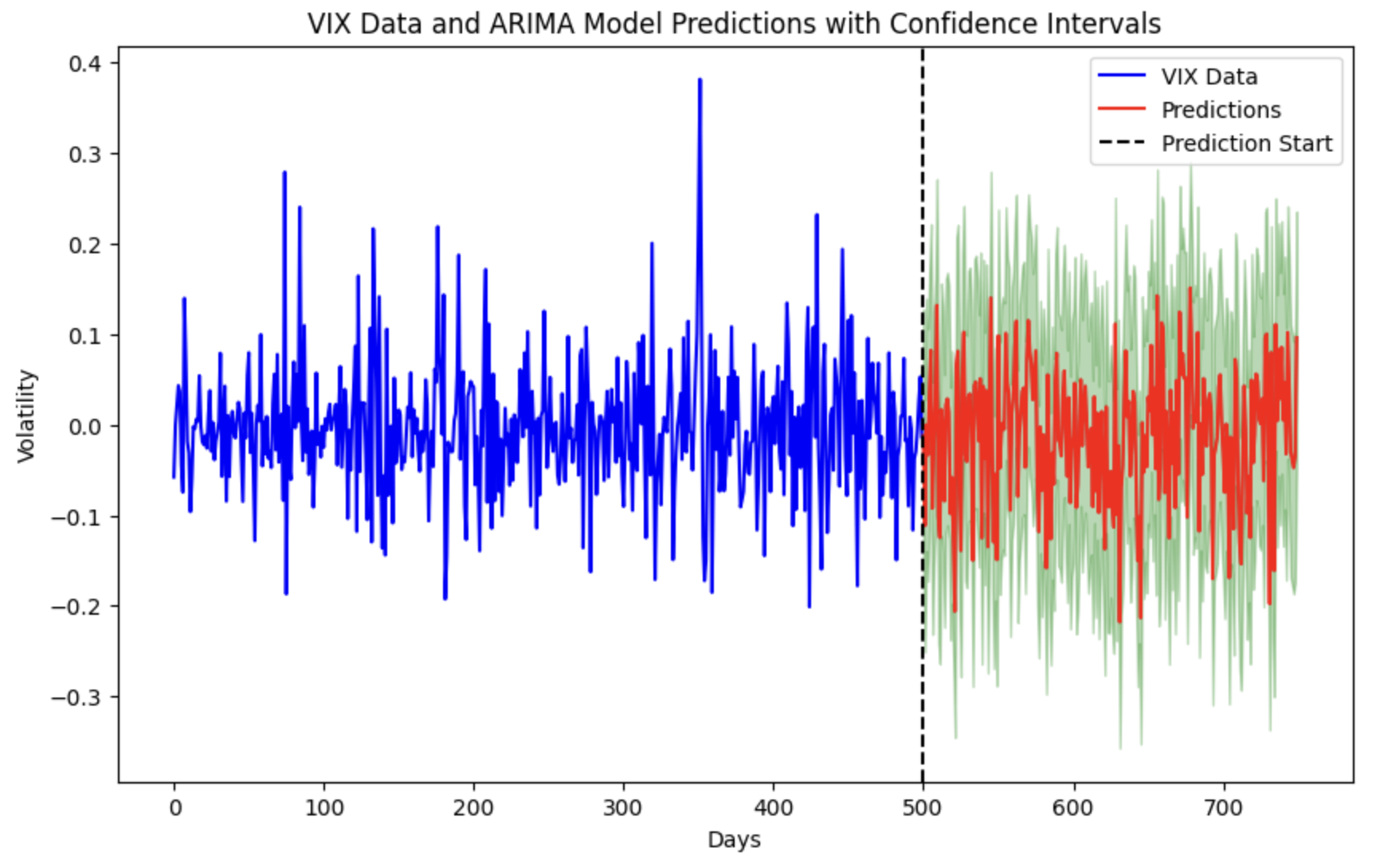}
        \caption{Prediction using a combination of the time series model(ARMA(2,1)) and GBM, where the green area is 95\% confidence interval}
        \label{fig:Prediction2}
    \end{minipage}
    \centering
\end{figure}

\begin{figure}[h!]
    \centering
    \begin{subfigure}[b]{0.45\textwidth}
        \centering
        \includegraphics[width=\linewidth]{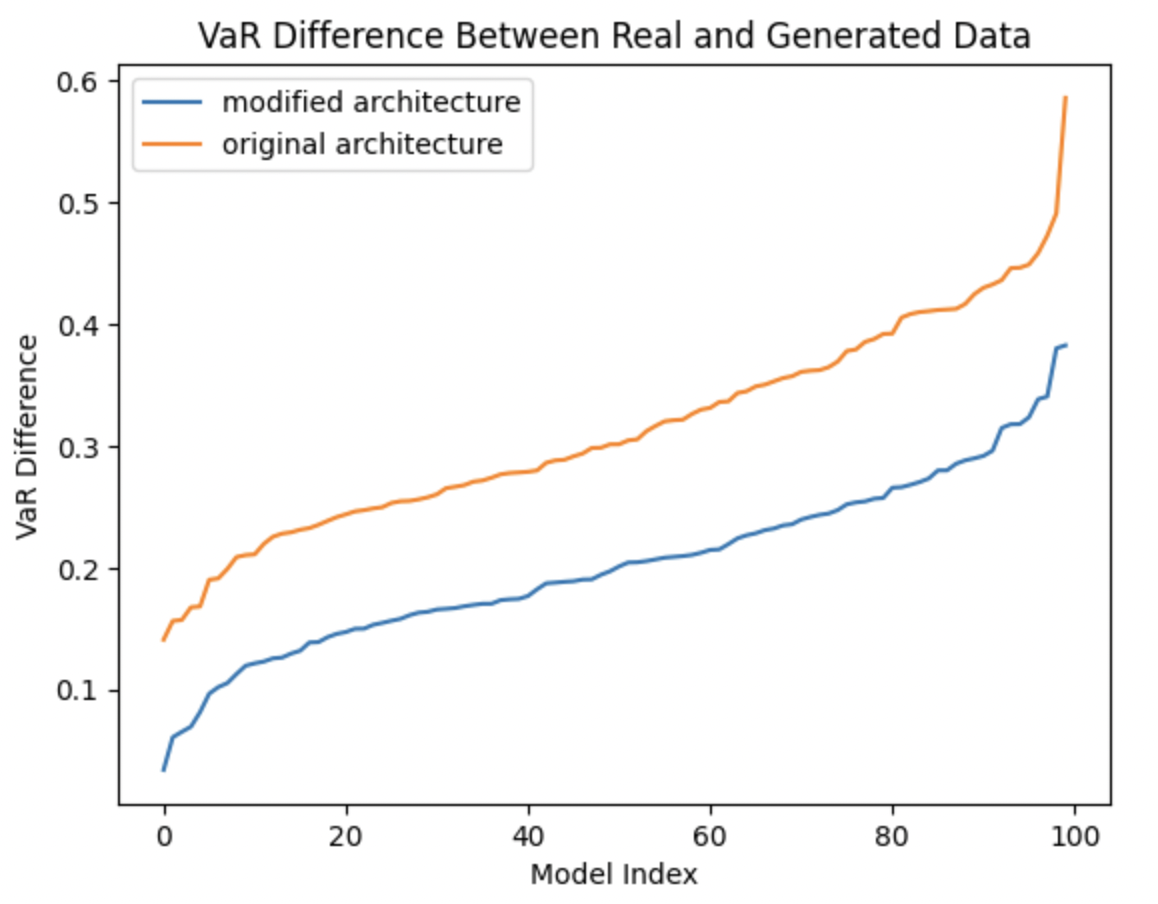}
        \caption{Comparison of VaR difference between the benchmark and the improved time series}
        \label{fig:VaR(origin vs improved time series)}
    \end{subfigure}
    \hfill
    \begin{subfigure}[b]{0.45\textwidth}
        \centering
        \includegraphics[width=\linewidth]{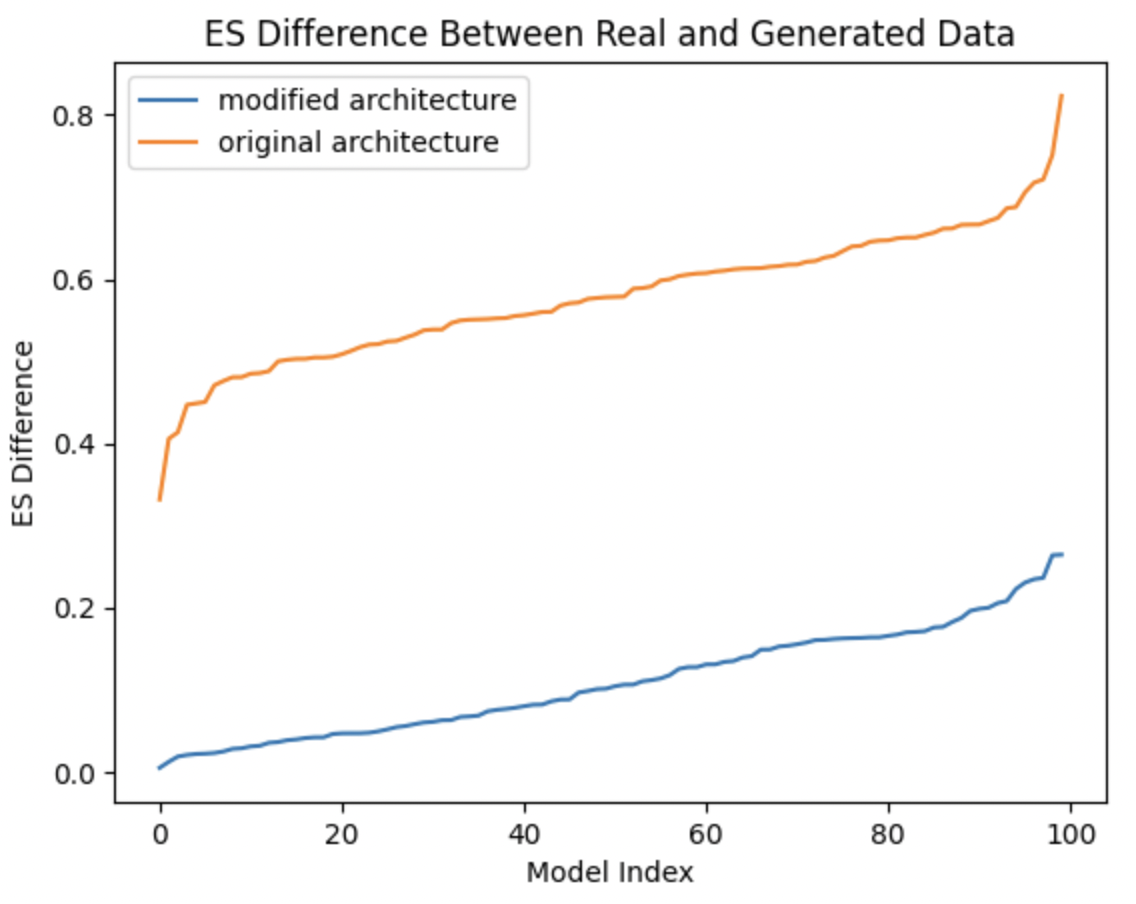}
        \caption{Comparison of ES difference between the benchmark and the improved time series}
        \label{fig:ES(origin vs improved time series)}
    \end{subfigure}
    
    \caption{Comparison of VaR and ES differences at 5\% level between the benchmark and the improved time series model}
\end{figure}

For consistency with Figure \ref{fig:Prediction}, we use the same data for training and prediction. The prediction data is decomposed as shown in Figure \ref{decomposition}. By substituting the noise component, we obtain an ideal prediction illustrated in Figure \ref{fig:Prediction2}. It is evident that this combined approach captures the volatility much more effectively than the model in Figure \ref{fig:Prediction}. This methodology will be utilized to generate the input sequence for the generator, and the results are shown in Figure \ref{fig:VaR(origin vs improved time series)} and \ref{fig:ES(origin vs improved time series)}. 

Similar to the performance observed with the time series model (see Figure \ref{fig:ES(origin vs time series)}), the improved time series model exhibits comparable performance in estimating ES, which aligns with our expectations given the trend and seasonal components retained in the model. However, the improved model shows a slight enhancement in VaR estimation. Specifically, with the improved model, 40 models fall under a VaR difference of 0.18, and 60 models are under 0.22, compared to 0.2 and 0.24 in the original time series model (see Figure \ref{fig:VaR(origin vs time series)}), respectively. Both models also have approximately 20 models under 0.14 and 80 models under 0.28, suggesting that the improved time series model offers slightly better performance.

Despite these improvements, the enhanced time series model still underperforms in VaR estimation compared to the simpler models that use historical data and mean-variance inputs. This discrepancy is challenging to explain given the current limitations in understanding neural network behavior. Further investigation could involve combining different models with varying weights or assigning a lower weight to the noise term in the time series model. However, due to the scope of this paper, further experiments on this issue are not conducted.

\section{Comparison between Tail-GAN and WGAN under FE-GAN}
\label{comp}

In this section, we implement Tail-GAN by modifying the loss and gradient functions, as specified earlier, and evaluate its performance against WGAN within the FE-GAN framework. Previous studies, such as \cite{cont2022tailgan}, demonstrated that Tail-GAN outperforms WGAN at smaller $\alpha$-quantiles while showing comparable performance at larger $\alpha$-quantiles. Building on this work, we extend the comparison by assessing both models under the FE-GAN architecture introduced in Section \ref{Modify}. Specifically, we compare their VaR and ES differences at the 5\% level across various input sequences, including historical data, sequences generated with mean and variance under the GBM assumption, a pure time series model, and an aggregated model combining time series and GBM inputs.

\subsection{Comparison Using Historical Data}

\begin{figure}[h!]
    \centering
    \begin{subfigure}[b]{0.45\textwidth}
        \centering
        \includegraphics[width=\linewidth]{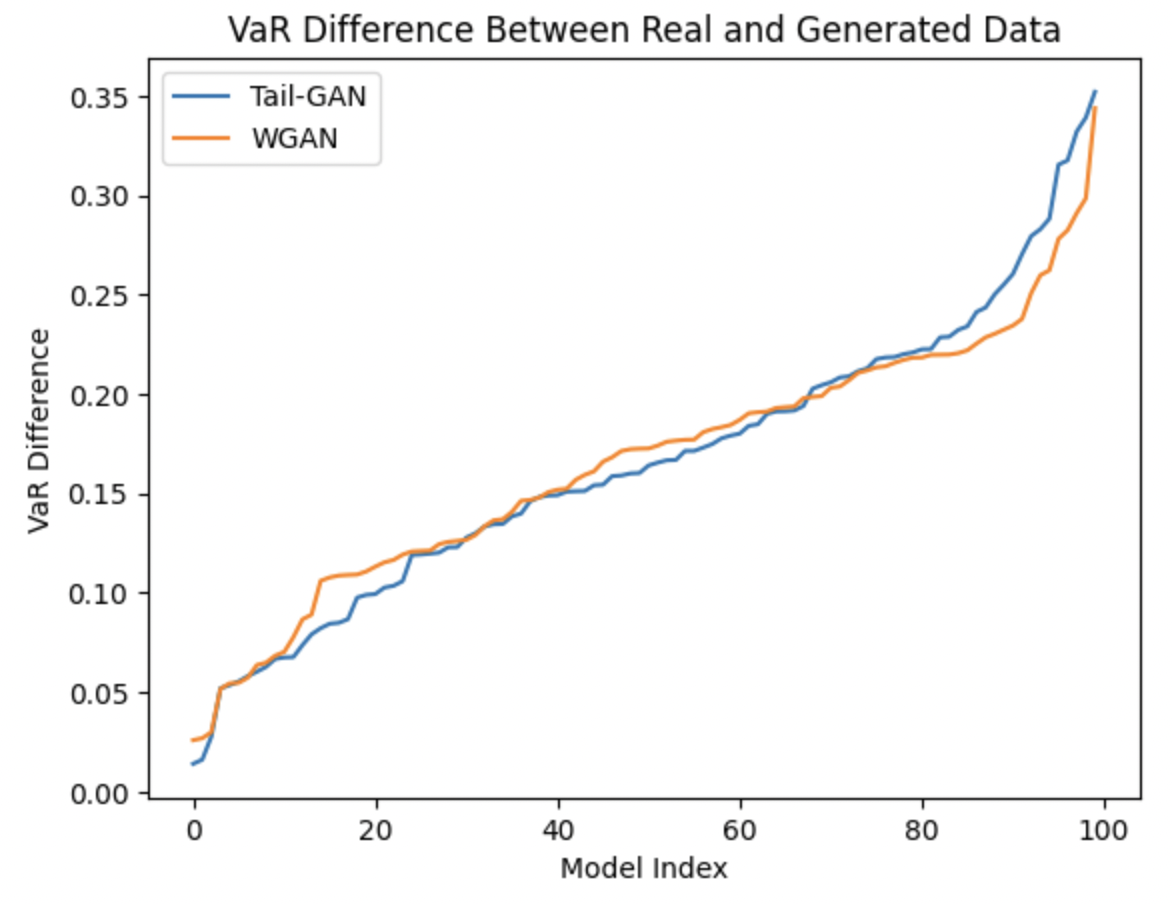}
        \caption{Comparison of VaR differences between WGAN and Tail-GAN under the FE-GAN architecture, using historical data as the input sequence.}
        \label{fig:VaR(hist vs hist+tail-GAN)}
    \end{subfigure}
    \hfill
    \begin{subfigure}[b]{0.45\textwidth}
        \centering
        \includegraphics[width=\linewidth]{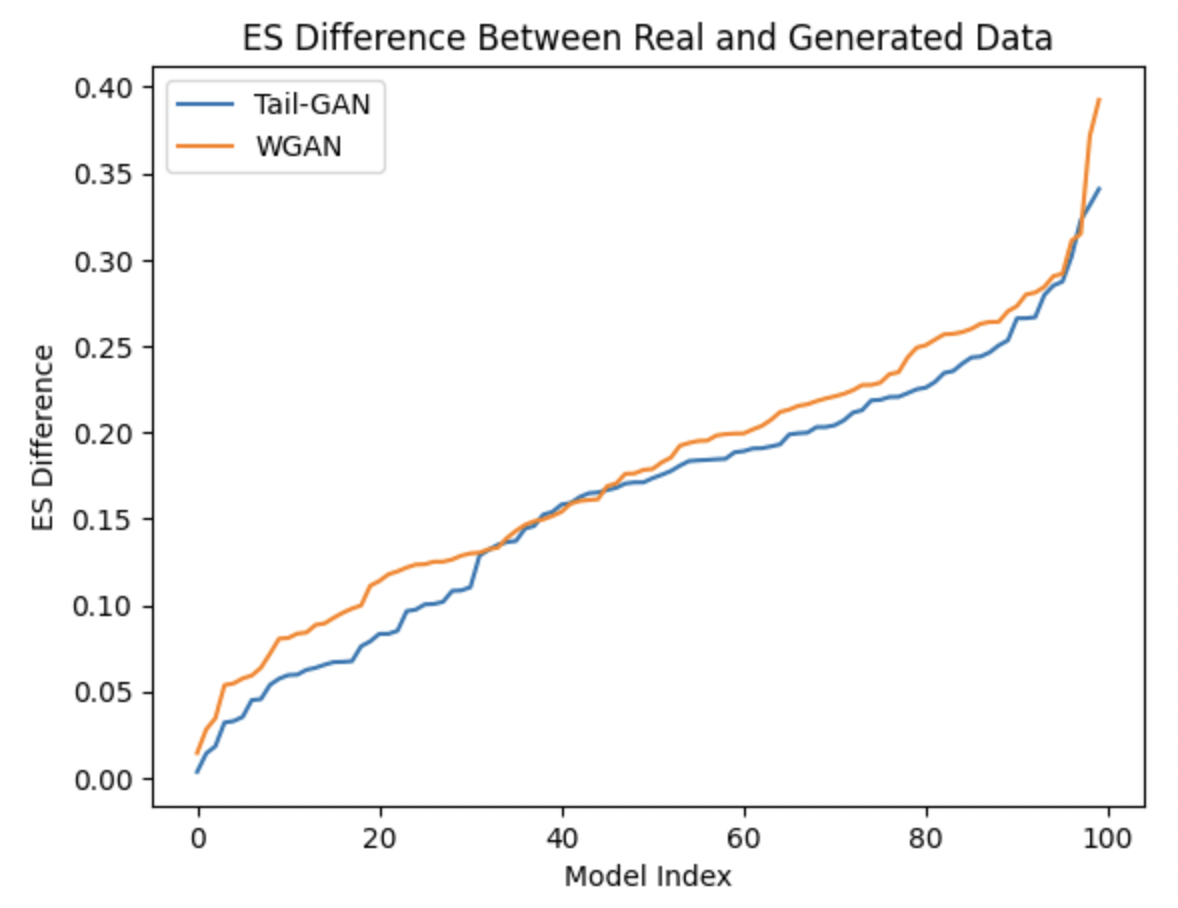}
        \caption{Comparison of ES differences between WGAN and Tail-GAN under the FE-GAN architecture, using historical data as the input sequence.}
        \label{fig:ES(hist vs hist+tail-GAN)}
    \end{subfigure}
    
    \caption{Comparison of VaR and ES differences at the 5\% level between WGAN and Tail-GAN under the FE-GAN architecture, using historical data as the input sequence.}
\end{figure}

Figures \ref{fig:VaR(hist vs hist+tail-GAN)} and \ref{fig:ES(hist vs hist+tail-GAN)} illustrate the performance of WGAN and Tail-GAN under the FE-GAN architecture, using historical data as the input sequence. In VaR estimation at the 5\% level (Figure \ref{fig:VaR(hist vs hist+tail-GAN)}), both models exhibit similar performance, with approximately 40 models achieving a difference below 0.15 and around 80 models under a difference of 0.22. However, for ES estimation (Figure \ref{fig:ES(hist vs hist+tail-GAN)}), Tail-GAN slightly outperforms WGAN. Specifically, Tail-GAN achieves 20 models with differences below 0.05 and 40 models under 0.13, compared to WGAN's 20 models under 0.13 and 40 models below 0.16.

These results indicate that Tail-GAN maintains its advantage over WGAN, particularly in ES estimation. The superior performance of Tail-GAN highlights its effectiveness in capturing tail dependencies and generating more accurate data under the FE-GAN framework, especially when the focus is on extreme quantiles.

\subsection{Comparison under the Assumption of GBM}

\begin{figure}[h!]
    \centering
    \begin{subfigure}[b]{0.45\textwidth}
        \centering
        \includegraphics[width=\linewidth]{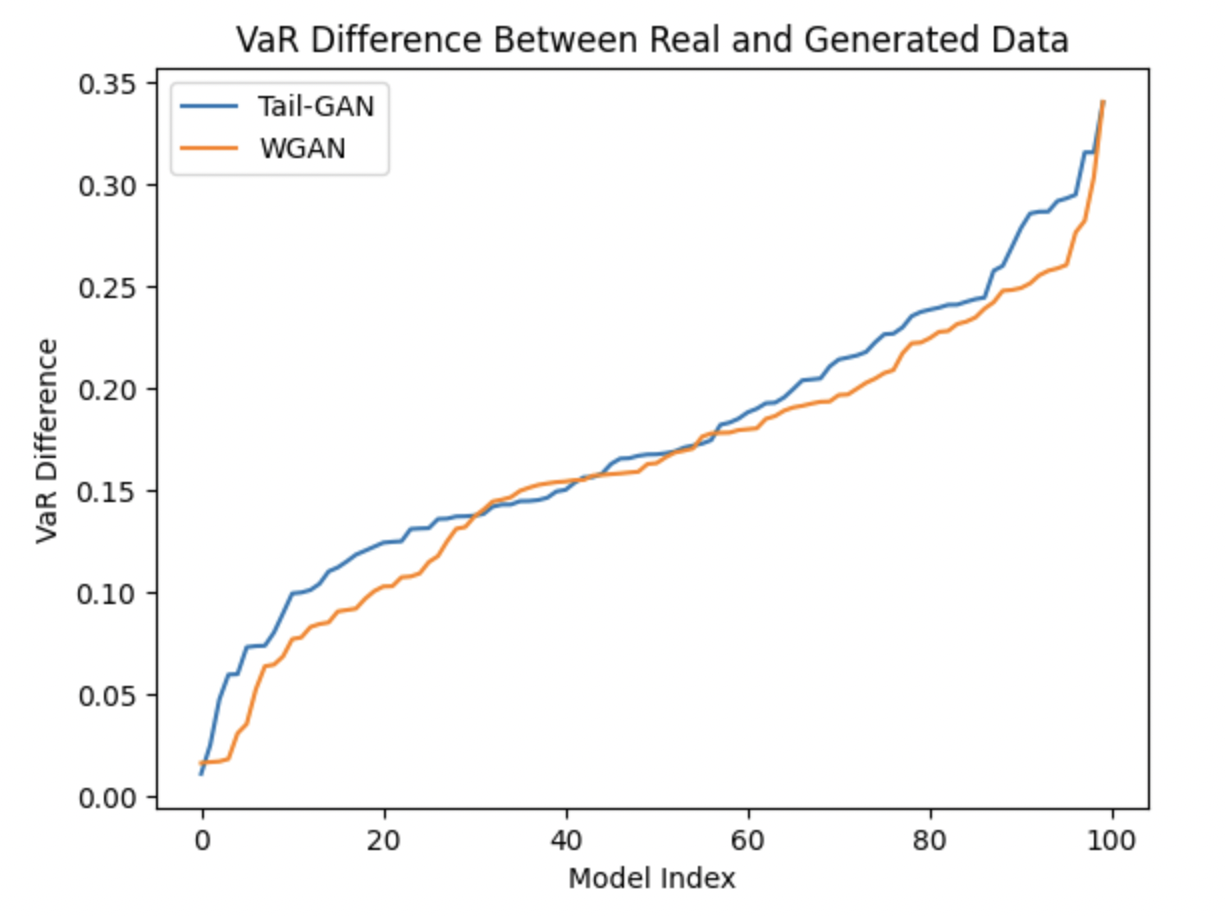}
        \caption{Comparison of VaR differences between WGAN and Tail-GAN under the FE-GAN architecture, using mean and variance to generate the input sequence (under the GBM assumption).}
        \label{fig:VaR(GBM vs GBM+tail-GAN)}
    \end{subfigure}
    \hfill
    \begin{subfigure}[b]{0.45\textwidth}
        \centering
        \includegraphics[width=\linewidth]{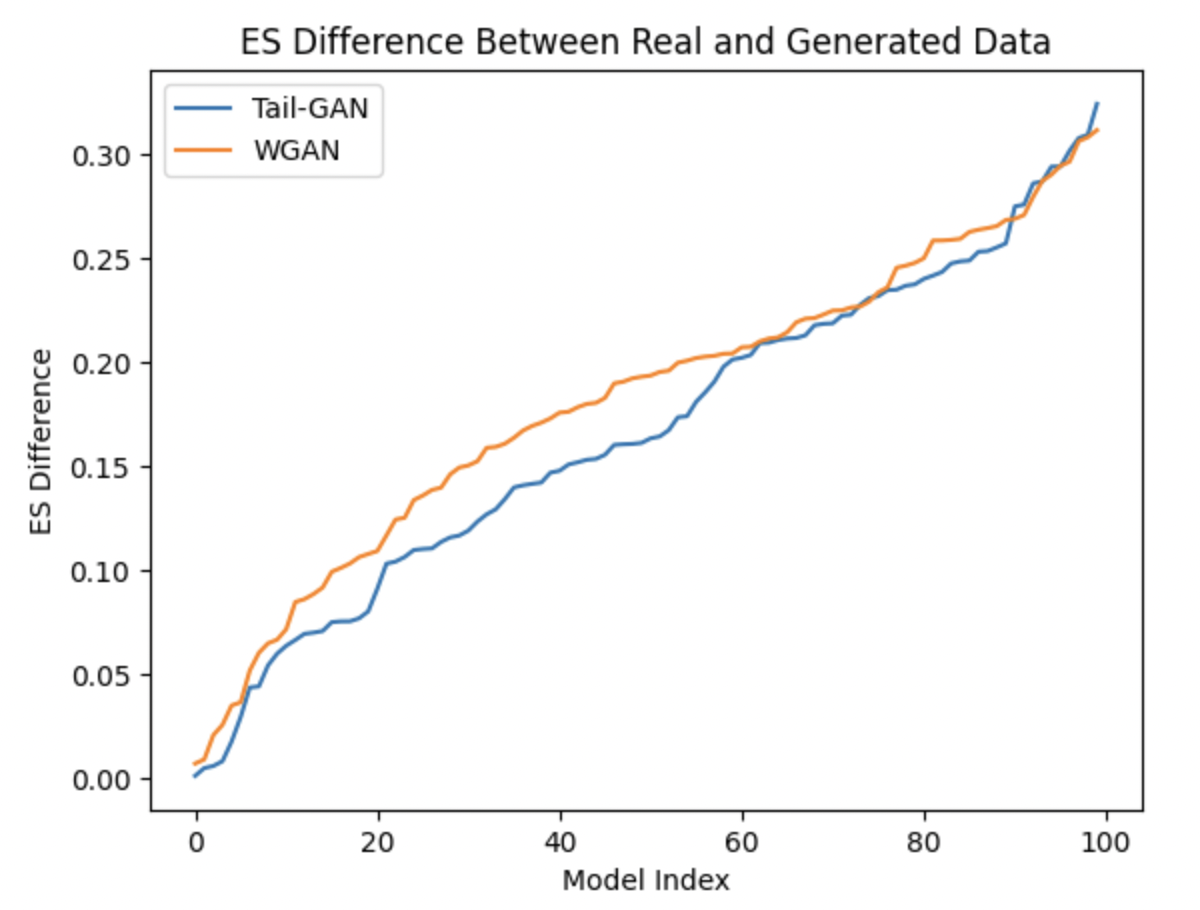}
        \caption{Comparison of ES differences between WGAN and Tail-GAN under the FE-GAN architecture, using mean and variance to generate the input sequence (under the GBM assumption).}
        \label{fig:ES(GBM vs GBM+tail-GAN)}
    \end{subfigure}
    
    \caption{Comparison of VaR and ES differences at the 5\% level between WGAN and Tail-GAN under the FE-GAN architecture, using mean and variance to generate the input sequence (under the GBM assumption).}
\end{figure}

Similar to the results with historical data, WGAN and Tail-GAN exhibit comparable performance in VaR estimation at the 5\% level, as shown in Figure \ref{fig:VaR(GBM vs GBM+tail-GAN)}. However, Tail-GAN demonstrates a marginal advantage in ES estimation, with 40 models achieving differences below 0.15 and 60 models below 0.18, compared to WGAN's 40 models below 0.17 and 60 models below 0.22, as depicted in Figure \ref{fig:ES(GBM vs GBM+tail-GAN)}. 

These results further confirm that Tail-GAN is better suited for estimating ES under the FE-GAN architecture, leveraging its design to optimize for joint elicitability of VaR and ES. While WGAN may capture broader data features, it does not achieve the same precision as Tail-GAN for risk measure estimation, particularly at extreme quantiles.

In practical applications, WGAN could be more appropriate if the objective extends beyond risk measure estimation, such as generating synthetic datasets for diverse financial analyses. However, for tasks focused on VaR and ES, particularly at smaller quantiles, Tail-GAN remains a more effective choice. Exploring alternative loss functions tailored for specific measures, such as VaR, could further enhance the accuracy of WGAN or Tail-GAN for task-specific applications, but this remains outside the scope of this paper.

\subsection{Comparison of Time Series Models}

\begin{figure}[h!]
    \centering
    \begin{subfigure}[b]{0.45\textwidth}
        \centering
        \includegraphics[width=\linewidth]{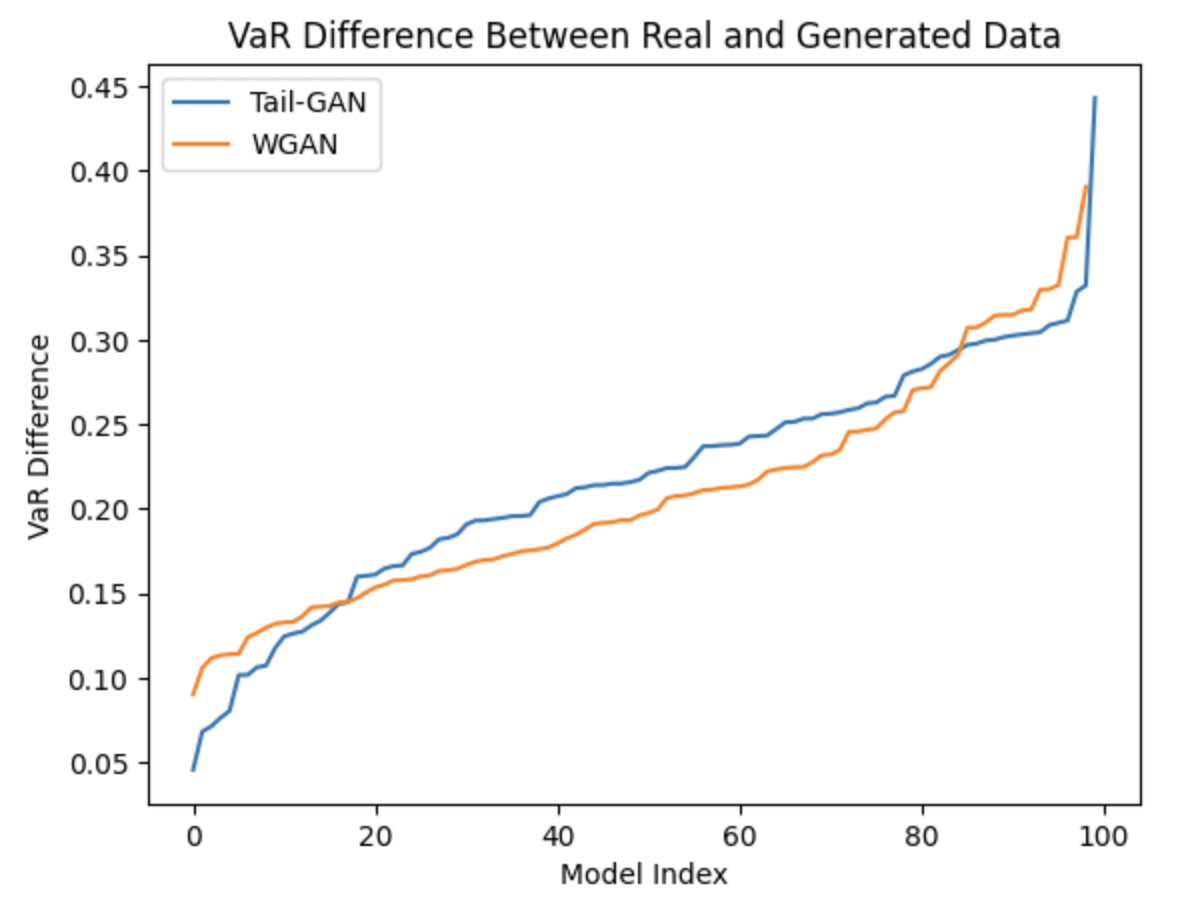}
        \caption{Comparison of VaR differences between WGAN and Tail-GAN under the FE-GAN architecture, using time series models to generate the input sequence.}
        \label{fig:VaR(time series vs time series+tail-GAN)}
    \end{subfigure}
    \hfill
    \begin{subfigure}[b]{0.45\textwidth}
        \centering
        \includegraphics[width=\linewidth]{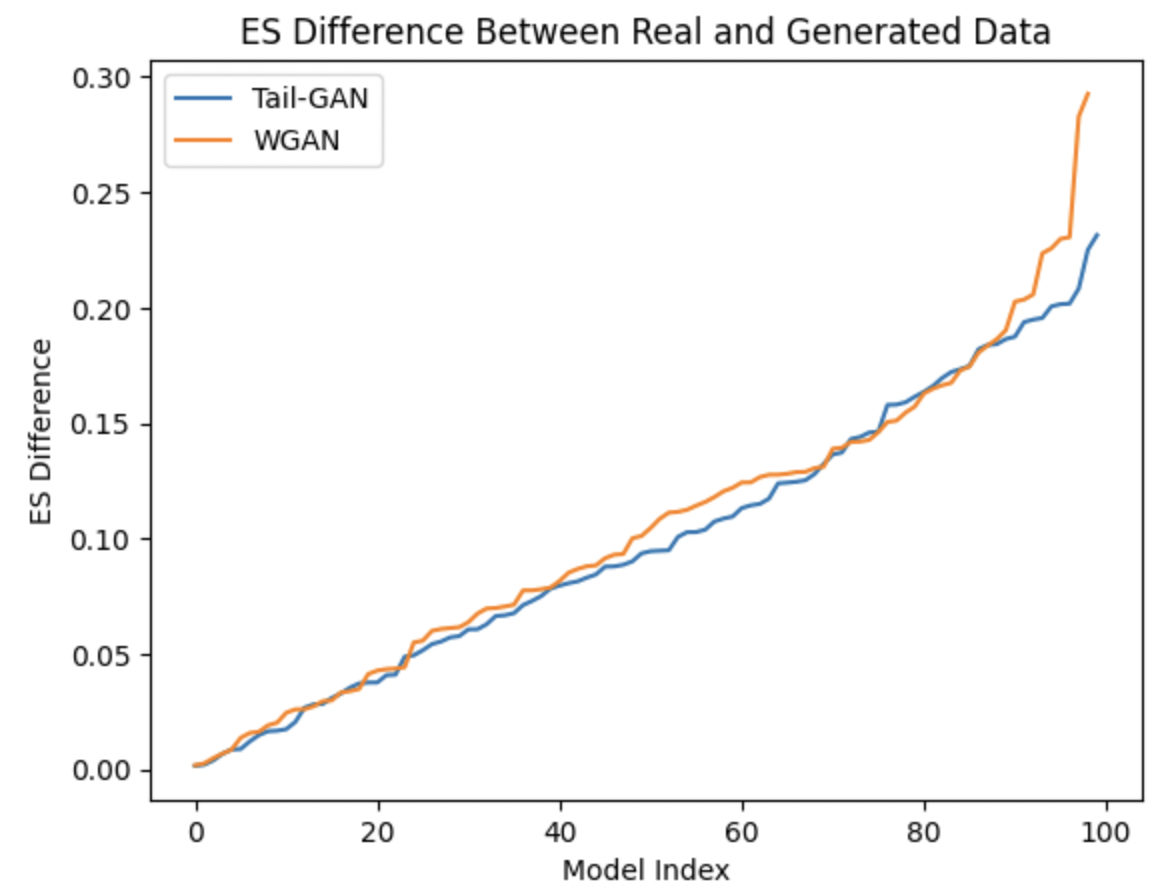}
        \caption{Comparison of ES differences between WGAN and Tail-GAN under the FE-GAN architecture, using time series models to generate the input sequence.}
        \label{fig:ES(time series vs time series+tail-GAN)}
    \end{subfigure}
    
    \caption{Comparison of VaR and ES differences at the 5\% level between WGAN and Tail-GAN under the FE-GAN architecture, using time series models to generate the input sequence.}
\end{figure}

When using time series models to generate input sequences, Tail-GAN does not show a significant improvement over WGAN in VaR estimation, as depicted in Figure \ref{fig:VaR(time series vs time series+tail-GAN)}. Both models underperform compared to simpler inputs such as historical data and mean-variance sequences (see Figures \ref{fig:VaR(hist vs hist+tail-GAN)} and \ref{fig:VaR(GBM vs GBM+tail-GAN)}), reaffirming the limitations of time series models in capturing volatility features.

For ES estimation, Figure \ref{fig:ES(time series vs time series+tail-GAN)} shows that Tail-GAN performs similarly to WGAN, which contrasts with its superior performance observed under simpler input sequences (see Figures \ref{fig:ES(hist vs hist+tail-GAN)} and \ref{fig:ES(GBM vs GBM+tail-GAN)}). This result aligns with the expectation that time series models, while effective in capturing trends, are less suited for estimating tail-dependent measures like ES.

These findings suggest that while time series models offer some advantages in trend prediction, their limitations in modeling volatility hinder their effectiveness for VaR and ES estimation under Tail-GAN. Further refinement, such as integrating additional volatility-focused components or hybrid approaches, may be required to improve their applicability for tail risk estimation.

\subsection{Comparison of Improved Time Series Models}

\begin{figure}[h!]
    \centering
    \begin{subfigure}[b]{0.45\textwidth}
        \centering
        \includegraphics[width=\linewidth]{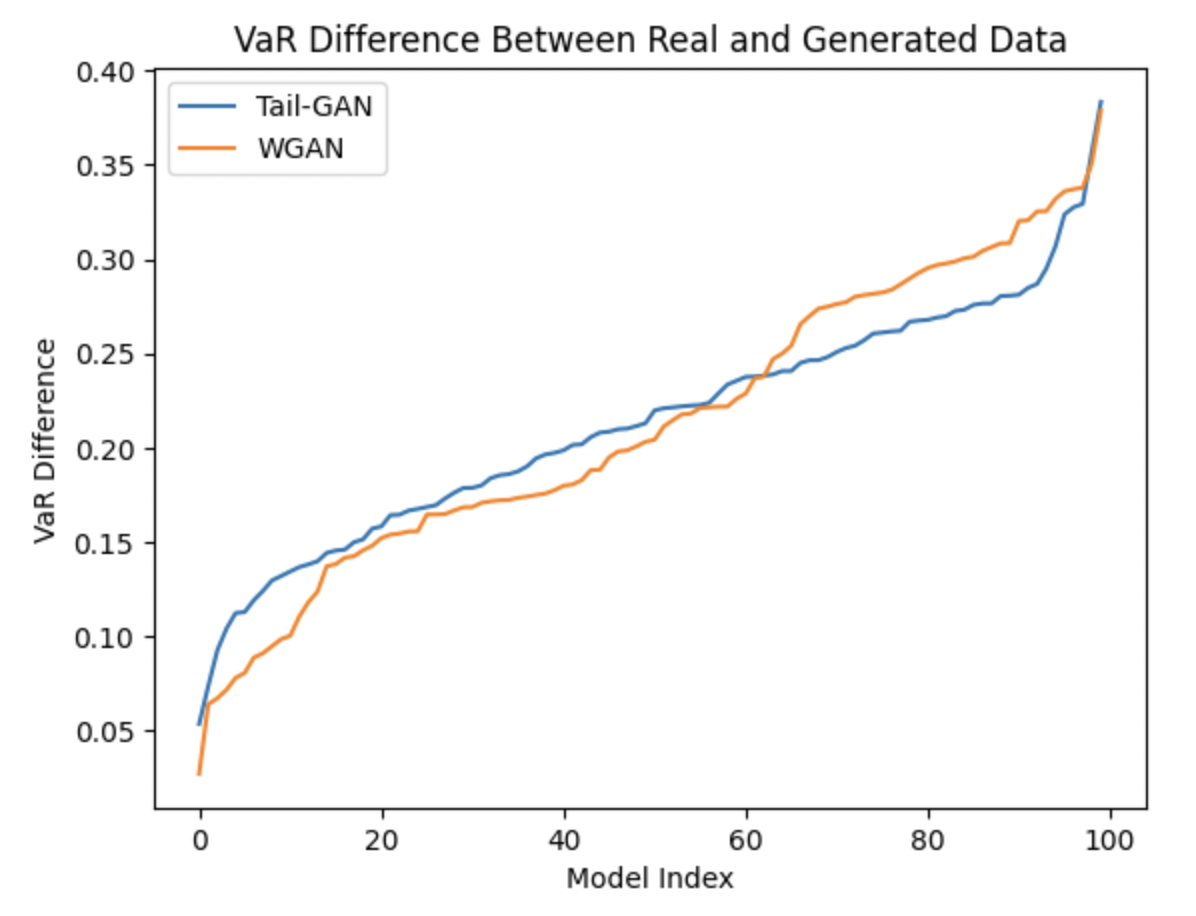}
        \caption{Comparison of VaR differences between WGAN and Tail-GAN under the FE-GAN architecture, using the improved time series model as the input sequence.}
        \label{fig:VaR(improved time series vs improved time series+tail-GAN)}
    \end{subfigure}
    \hfill
    \begin{subfigure}[b]{0.45\textwidth}
        \centering
        \includegraphics[width=\linewidth]{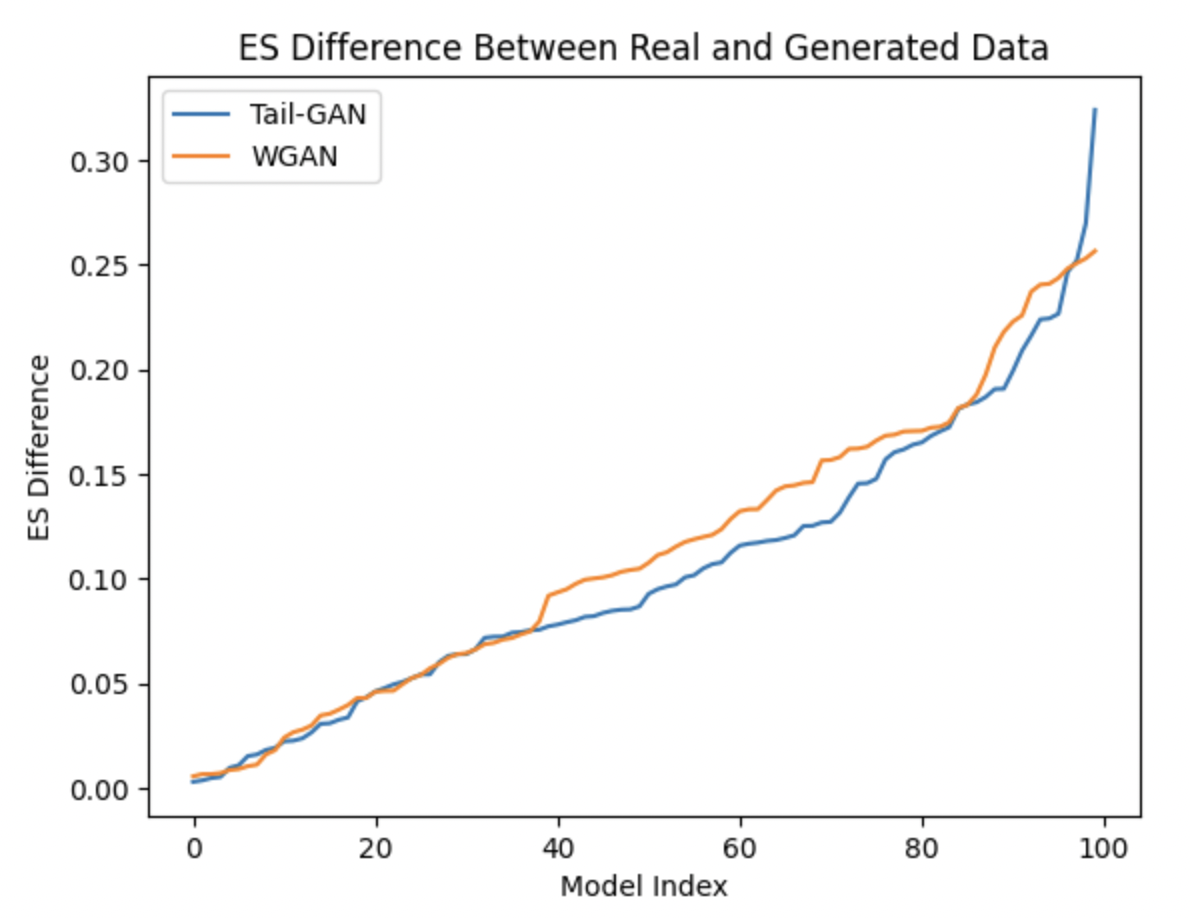}
        \caption{Comparison of ES differences between WGAN and Tail-GAN under the FE-GAN architecture, using the improved time series model as the input sequence.}
        \label{fig:ES(improved time series vs improved time series+tail-GAN)}
    \end{subfigure}
    \caption{Comparison of VaR and ES differences at the 5\% level between WGAN and Tail-GAN under the FE-GAN architecture, using the improved time series model as the input sequence.}
    \label{fig:VaR_ES_Comparison}
\end{figure}

Figures \ref{fig:VaR(improved time series vs improved time series+tail-GAN)} and \ref{fig:ES(improved time series vs improved time series+tail-GAN)} show that, even with the modifications in the time series model, Tail-GAN and WGAN exhibit similar performance in estimating both VaR and ES at the 5\% level. This suggests that the improved time series model reduces the performance gap between WGAN and Tail-GAN for these risk measures. 

One possible explanation is that the enhanced structure of the time series model introduces additional trend and seasonal information, which improves the baseline performance of both WGAN and Tail-GAN. Consequently, the advantages of Tail-GAN’s tailored loss function in capturing extreme quantiles become less pronounced when combined with an input sequence that already captures much of the underlying data structure.

While these results highlight the potential of improved time series models to enhance the overall estimation process, further investigation is needed to understand why Tail-GAN does not show significant improvement over WGAN in this context. Future work could explore whether alternative loss functions or a reweighted emphasis on the volatility component could yield a larger performance differential between the two models.

\section{Discussion}

\subsection{Reflection on Research Objectives}
This study explored the use of FE-GAN for financial risk management, specifically for estimating VaR and ES. FE-GAN enhances existing GANs architectures by incorporating an additional input sequence derived from historical data, GBM, or time series models. This enhancement aims to improve the generator's ability to capture temporal patterns in financial data, especially under conditions of volatility and extreme events. The results indicate that FE-GAN achieves significant improvements in training efficiency and risk measure estimation compared to traditional WGAN.

\subsection{Key Findings and Comparative Analysis}
The key finding of this study is that FE-GAN consistently outperforms traditional WGAN in estimating VaR and ES by leveraging enriched input sequences. The use of historical data and mean-variance inputs derived under the GBM assumption demonstrated superior performance in VaR estimation, while time series-based input sequences proved more effective for ES estimation. Combining time series components (trend and seasonality) with volatility modeling under GBM further enhanced the model's ability to estimate VaR, demonstrating a complementary relationship between these approaches.

In the comparative analysis, Tail-GAN under the FE-GAN framework outperformed WGAN in ES estimation across all input scenarios, aligning with findings from \cite{cont2022tailgan}. Tail-GAN's advantage stems from its loss function tailored for joint elicitability of VaR and ES, which enables better modeling of tail risks and extreme financial events. These results highlight the potential of FE-GAN to improve task-specific performance for financial risk measures.

\subsection{Interpretation and Speculation}
The observed improvements can be attributed to the enriched input sequence, which provides the generator with additional contextual information. This enhancement allows the generator to start closer to the true data distribution, accelerating convergence and improving estimation accuracy. By incorporating volatility features from GBM and trend/seasonality components from time series analysis, FE-GAN leverages complementary strengths to address the limitations of traditional GANs. This approach mirrors transfer learning, where pretraining with relevant information accelerates learning and enhances performance \cite{goodfellow2016deep}.

However, the results also revealed that time series models alone did not significantly improve VaR estimation. This may be due to the inherent difficulty in modeling volatility patterns. The integration of multiple input sources, as demonstrated in the hybrid approach, appears to be crucial for improving overall performance. Future work could explore dynamic weighting of input components to better balance the contributions of trend and volatility features.

Due to the inherent complexity and lack of explainability in neural networks, several intriguing results from this study remain difficult to interpret. Firstly, although WGAN and Tail-GAN both aim to estimate risk measures, it is unclear why Tail-GAN consistently outperforms WGAN in FE-GAN after training. Secondly, despite combining the strengths of the GBM and time series models, the improved time series model still exhibits inferior performance in VaR estimation compared to the simpler GBM model. Lastly, Tail-GAN shows superior performance when using historical data and the GBM model, yet this advantage is not observed when time series-based models are employed. These questions highlight the need for further research to better understand the underlying mechanisms at play.

\subsection{Study Limitations}
Several limitations must be acknowledged. First, FE-GAN relies on enriched input sequences, which require continuous and highly correlated temporal data. This restricts its applicability to domains like financial time series and limits its utility in areas. Second, the study focused exclusively on VaR and ES estimation using VIX data. While the results are promising within this context, the generalizability of FE-GAN to other datasets and risk measures remains untested.

Additionally, the reliance on fixed input generation methods, such as the 250-day trading window, may not be optimal for all forecasting scenarios. Shorter or dynamically adjusted windows might yield better results for specific use cases. Furthermore, the architecture and hyperparameters were not extensively tuned, which leaves room for further optimization \cite{zhang2020interpreting}. These factors highlight the need for broader exploration to refine the approach.

\subsection{Implications and Future Research Directions}
The findings of this study underscore the potential of FE-GAN for financial risk management. By enhancing existing GANs architectures with enriched input sequences, FE-GAN provides a flexible and effective tool for estimating VaR, ES, and potentially other risk measures. Future research could extend this approach to incorporate additional forecasting techniques, such as Long Short-Term Memory (LSTM) networks \cite{hochreiter1997long}, to improve temporal dependency modeling.

Moreover, alternative neural network configurations could be explored to enhance performance further. For example, integrating Conditional GANs (CGANs) \cite{mirza2014conditional} or Deep Convolutional GANs (DCGANs) \cite{radford2015unsupervised} may provide better capabilities for capturing complex patterns in financial data. Incorporating techniques like dropout \cite{srivastava2014dropout} or auxiliary networks could stabilize training and reduce the risk of overfitting or mode collapse.

Finally, the development of robust evaluation metrics, including confidence intervals for VaR and ES, could ensure that the model's outputs are reliable and not overly sensitive to specific datasets. By addressing these areas, future research could refine FE-GAN into a more versatile and robust tool for financial applications.

\section{Conclusion}

This study investigated the application of FE-GAN in financial risk management, focusing on improving the estimation of VaR and ES. By enhancing existing GANs architectures with additional input sequences, the study aimed to provide richer contextual information to the generator, thereby improving performance in risk estimation.

The findings demonstrate that FE-GAN significantly outperforms traditional WGAN in both VaR and ES estimation. Using historical data or mean-variance inputs derived under the GBM assumption, FE-GAN achieved notable reductions in VaR estimation errors. Similarly, the incorporation of time series components enhanced ES estimation, reflecting the complementary strengths of trend and seasonality analysis alongside volatility modeling. The hybrid approach, combining the volatility-capturing GBM model with time series-derived components, further improved VaR estimation, showcasing the benefits of integrating diverse input features.

Tail-GAN consistently outperformed WGAN under the FE-GAN framework, particularly in ES estimation, reaffirming its effectiveness in capturing tail risks. This result highlights the importance of task-specific architectures in addressing extreme quantile-based measures, even within an enhanced GANs setting.

Despite these promising results, several limitations remain. The reliance on highly correlated temporal data, such as financial time series, restricts the broader applicability of FE-GAN to other domains. Additionally, the study focused exclusively on VaR and ES, leaving the performance of FE-GAN on other risk measures unexplored. The fixed 250-day trading window, while suitable for this analysis, may not optimize performance for long-term forecasting or datasets with different temporal characteristics.

Future research could address these limitations by experimenting with alternative neural network architectures, exploring dynamic input generation methods, and testing the framework on other datasets and risk measures. Incorporating advanced temporal modeling techniques, such as LSTM networks or attention-based mechanisms, could further enhance FE-GAN's ability to capture complex temporal dependencies. These efforts would expand the utility of FE-GAN, not only within financial risk management but also in other domains requiring robust and accurate risk estimation.

\end{document}